\documentclass[journal,draftcls,onecolumn,12pt,twoside]{IEEEtran}
\normalsize
\IEEEoverridecommandlockouts

\usepackage{amsmath, amssymb, cite}
\usepackage{amsthm,bm,bbm}
\usepackage{mathtools}
\usepackage[small]{caption}
\usepackage{amsthm,multirow,color,amsfonts}
\usepackage{tabulary}
\usepackage{subfigure}
\usepackage{graphicx}
\usepackage{setspace}
\usepackage{enumerate}
\usepackage[]{algorithm2e}
\usepackage{comment}
\usepackage{hyperref}
\usepackage{cleveref}
		
\newcommand*\xor{\oplus}

\newtheorem{theo}{Theorem}

\newcommand\norm[1]{\left\lVert#1\right\rVert}

\newcommand{\mathb}{\mathbf}

\def\compactify{\itemsep=0pt \topsep=0pt \partopsep=0pt \parsep=0pt}
\let\latexusecounter=\usecounter

\IEEEaftertitletext{\vspace{-2\baselineskip}}

\begin{document}
\title{Distributed Computation over MAC via Kolmogorov-Arnold Representation}
\author{Derya Malak and Ali Tajer
\thanks{Derya Malak and Ali Tajer are with the department of Electrical, Computer, and Systems Engineering, Rensselaer Polytechnic Institute, Troy, NY 12180, USA (malakd@rpi.edu, tajer@ecse.rpi.edu).}
}

\maketitle

\begin{abstract}

Kolmogorov's representation theorem provides a framework for decomposing any arbitrary real-valued, multivariate, and continuous function into a two-layer nested superposition of a finite number of functions. The functions at these two layers, are referred to as the inner and outer functions with the key property that the design of the inner functions is independent of that of the original function of interest to be computed. This brings modularity and universality to the design of the inner function, and subsequently, a part of computation. This paper capitalizes on such modularity and universality in functional representation to propose two frameworks for distributed computation over the additive multiple access channels (MACs). In the first framework, each source encodes the inner representations and sends them over the additive MAC. Subsequently, the receiver computes the outer functions to compute the function of interest. Transmitting the values of the inner functions instead of the messages directly leads to compression gains. In the second approach, in order to further increase the compression rate,  the framework aims to also bring computing the outer functions to the source sites. Specifically, each source employs a graph-coloring-based approach to perform joint functional compression of the inner and the outer functions, which may attain further compression savings over the former. These modular encoding schemes provide an exact representation in the asymptotic regime and the non-asymptotic regime. Contrasting these with the baseline model where sources directly transmit data over MAC, we observe gains. To showcase the gains of these two frameworks and their discrepancies, they are applied to a number of commonly used computations in distributed systems, e.g., computing products, $\ell_m$-norms, polynomial functions, extremum values of functions, and affine transformations.


\end{abstract}

\section{Introduction}
\label{section:introduction}

Communication networks are being increasingly deployed for computation purposes. Examples include over-the-air function computation in wireless networks \cite{abari2016over}, private-key encryption \cite{komargodski2020minicrypt},  
communication load versus distributed computation latency trade-offs \cite{ezzeldin2017communication,li2017fundamental,li2017coding}, as well as storage, computation, communication trade-offs \cite{yan2018storage,yan2019storage}, coded matrix multiplication \cite{YuAliAve2017,lee2017high} accounting for stragglers and sparsity \cite{ramamoorthy2019universally}, and the heterogeneity of the clusters \cite{reisizadeh2019coded}, and coded distributed convolution with stragglers within a deadline \cite{dutta2017coded}, which is beyond distributed matrix multiplication, among many others. When the objective is to retried a function of the source data rather than the data, integration computation and communication can provide significant saving in communication resources due to the often lower information content of the functions compared to that of the data. Furthermore, in privacy-constrained  networks, communicating computed functional values facilitates protecting private source data. Motivated by these factors, in this paper, we consider the problem of distributed compression of data for recovering a function at the receiver.

In this paper, we consider communication over the multiple access channel, in which the receiver is interested in computing a function of the information sources. We provide a systematic approach that breaks an arbitrary computation task into two distinct sets of computation tasks performed by the sources and receiver. This framework allows for savings in communication in two ways. First, the sources carry out a part of computation and communicate functionals of the data instead of the data. Secondly, what is computed and communicated by the sources is designed with the perspective of the needs of the computation sub-task at the receiver for recovering the function of interest. This framework emerges from 
a general theory of decomposing multivariate continuous functions developed by Kolmogorov and Arnold~\cite{kolmogorov1957representation}. This theory inspires creating a distributed computation framework via identifying possible decomposition methods for distributed encoding and transmission of signals over an additive multiple access channel and recovering the functional representation at a common destination. From an information-theoretic perspective, such functional decomposition translates the function computation tasks to distributed source compression and functional compression.

\subsection{Related Work}
\label{setion:related}

Distributed source coding is a direct, albeit not effective, approach to distributed computation.  The Slepian--Wolf coding provides information-theoretic limits lower bounds on the coding rate in order to  recover $n$ sources  $(X_1,\dots, X_n)$ at a receiver with an arbitrarily small error probability. Specifically, by denoting the entropy of random variable $X_p$ by $H(X_p)$ and its lossless coding rate by $X_p$  we have~\cite{slepian1973noiseless}
\begin{align}
\sum\limits_{p\in\mathcal{S}}R_p\geq H\Big((X_p)_{p\in\mathcal{S}}\vert \{X_p\}_{p\notin\mathcal{S}}\Big)\ ,\quad \forall \mathcal{S}\subseteq \{1,\hdots, n\}\ .
\end{align}
While allowing for some compression due to correlation among sources, and subsequently, saving in communication resources, such a compression-based approach does not capture and account for the additional inherent redundancies associated with computation. Motivated by this, there have been extensive recent studies on distributed compression for computing functions of the sources. Some examples include compression for multiple descriptions of functions \cite{gamal1982achievable}, special functions such as addition \cite{korner1979encode}, multiplication with side information 
\cite{watanabe2013rate}, and the graph-entropic approaches \cite{alon1996source,orlitsky2006coding,korner1973coding, doshi2010functional,doshi2007source,feizi2014network,sefidgaran2011computing}. There also exist approaches that focus on the distributed computation of special classes of functions in networks. For example, the study in~\cite{mosk2006computing} has developed a gossip-based algorithm for computing separable functions. 
In the sensor network literature, over-the-air computation has been widely studied leveraging the superposition property of the MAC. Several examples include the investigation of the maximum rate of computing and communicating symmetric functions in~\cite{giridhar2005computing}, the computation rate for approximating nomographic functions in~\cite{frey2021over},  and 
derivation of the mean-squared error (MSE) of 
over-the-air computation of the linear superposition of spatially and temporally correlated signals in \cite{liu2021over}. Furthermore, for any algorithm used by the nodes to communicate and compute, a fundamental lower bound on distributed computation to achieve the desired MSE criterion is derived in \cite{ayaso2010information}. In \cite{halbawi2018improving} authors study an online distributed coded algorithm to compute the gradient from the smallest possible number of machines for heavy-tailed delays. 
Other approaches focus on generalizing the Slepian-Wolf compression and compressed sensing results to distributed functional compressed sensing \cite{muthukrishnan2006some}, and compressed sensing of functions over networks \cite{feizi2010compressive}.

In wireless networks, computing over the multiple access channel has received extensive attention due to 
practical scenarios including sensor networks \cite{giridhar2005computing}, over-the-air computing \cite{abari2016over}, identification of the differences between sources \cite{korner1979encode}, and semantics-aware radio access technologies \cite{dommel2021joint}, or frameworks that require the distributed computation of functions via exploiting the structural match between the channel and the sum of observations \cite{nazer2007computation}. Furthermore, computation coding can achieve the multicast capacity of linear MACs \cite{nazer2007computation}. Some representative studies pertaining to computation over MAC include 
distributed joint source-channel coding \cite{rajesh2007source}, coding for computing with side information \cite{rajesh2008distributed,varshneya2006distributed}, and the reconstruction of the sources over Gaussian MAC subject to, e.g., a mean-squared error distortion criterion \cite{lapidoth2011communicating}, an average-power constrained MAC \cite{lapidoth2010sending}, and 
constraints on the powers of the received signals \cite{gastpar2004gaussian}. 
K\"orner and Marton have shown the effectiveness of coding for function decoding instead of decoding the full information \cite{korner1979encode}.
The study in~\cite{zhu2018communication} establishes duality results for the codes that can be leveraged both for computation and compression over MAC. 
The study in \cite{padakandla2013computing} devises nested codes for computation summations over MAC, and the framework in \cite{lim2017towards} applies linear nested codes to improve the achievable compute–forward region for MAC. In \cite{nazer2007computation}, authors investigate the fundamental limits of coding for computation of linear functions over MAC.  
The scheme in~\cite{ray2006separation} demonstrates the robustness of source-channel separation for noisy MAC when noise is independent of the inputs is.  
The study in~\cite{wagner2008rate} analyzes the rate region of the vector-valued Gaussian two-encoder source-coding problem and shows that a simple architecture that separates the analog and digital aspects of the compression is optimal. On the other hand, in some cases, the best achievable distortion in a network setting is achieved via joint source-channel coding \cite{gastpar2005power}. 
There also exist other studies that  devise approaches to source-channel matching schemes, demonstrating the suboptimality of separation-based schemes for computation over MAC~\cite{nazer2007computation ,cover1980multiple, ahlswede1983source, gunduz2009source}.

\subsection{Contributions}
\label{setion:contributions}

Inspired by the general theory for KA representation, which provides a decomposition for any real-valued and multivariate continuous function, we provide a framework for distributed computation of real-valued functions over the additive multiple access channels.  We consider zero-error compression, where we seek an exact representation at the receiver.  The key properties of the KA representation that facilitate such distributed computation are that (i)~computing a multivariate function can be decomposed into computing a nested set of inner and outer functions, and (ii)~the design of the inner functions is independent of the choice of the multivariate function of interest to be computed.

Besides a baseline model that involves direct transmission over the MAC and furnishes a benchmark for performance evaluation, we propose two compression approaches to the distributed computation of a function $f:\mathbb{R}^n\rightarrow\mathbb{R}$ in an $n$-user MAC by leveraging the structures of the inner and outer functions in the KA representation of $f$.

\begin{enumerate}
\item {\bf Compressing the inner functions:}  In this approach, only the inner layer of the KA representation of $f$ is leveraged by the sources. This allows sources to perform compression by transmitting only the functional representation of their messages necessary for computing $f$ at the receiver. In this approach, the additive nature of the channel superimposes the ingredients computed by different sources, and the receiver applies the outer functions to that channel output. Hence, in this approach, the inner and outer function computation is split between the sources and the receiver, respectively. This modular encoding scheme provides an exact representation both in the asymptotic regime as well as even in the non-asymptotic regime.

\item  {\bf Compressing the inner \& outer functions:} In the second approach, we further leverage the structure of the outer layers of the KA representation of $f$ at the sources to enhance the compression rate. To facilitate this, each source employs a graph coloring approach to capture the information-theoretic limits of compression. In the coloring-based scheme, each source creates a (set of) characteristic graph(s) and then efficiently encodes them before transmission. Subsequently, the receiver decodes the colors instead of evaluating the outer function representation. The additional color encoding operation can be modularized, and it provides additional saving in compression over the scheme that compresses only the inner functions \footnote{Due to the distributed nature of the problem, each source builds the characteristic graphs and their colorings independently, i.e., without observing the realizations of the other sources.}. The additional compression gains are viable due to exploiting the structure of the function at the encoder, then grouping the set of outcomes that needs to be distinguished at the receiver by assigning each set a unique color to prevent confusion at the receiver. This approach can ensure asymptotically zero-error computation of the function. In the non-asymptotic regime, the coloring approach can provide a quantized description of the function by encoding sufficiently large power graphs of the characteristic graphs.

\end{enumerate}

We characterize and contrast the gains of these distributed compression models for computing a set of commonly-used functions beyond requiring linearity such as multiplication, $\ell_m$-norm, polynomial, extremum, and affine transformations over Gaussian MAC. Exploiting the universality of inner functions, these functions can be used as baselines for computing a broader class of functions.

The paper is organized as follows. In Sect. \ref{section:preliminaries} we detail the KA representation theorem and its implications on the distributed computation of data over MAC. Then, in Sect. \ref{section:inner}, leveraging the KA representation and the universality of the inner functions, we devise a modular distributed computing framework for zero-error computing via compressing the inner functions. Next, in Sect. \ref{section:innerandouter}, exploiting a graph-coloring approach and encoding characteristic graphs of the sources, we provide a joint functional compression at the source site. Through Sects. \ref{section:inner} and \ref{section:innerandouter} we contrast the savings of different compression schemes for the functions outlined above. Finally, in Sect. \ref{section:extensions}, we provide some concluding remarks and future directions.

\section{Preliminaries}
\label{section:preliminaries}

\subsection{Kolmogorov-Arnold Representation}
\label{section:KA}


The KA representation theorem states that any real-valued,  multivariate continuous function $f:\mathbb{R}^n\rightarrow\mathbb{R}$ can be described as the superposition of $2n+1$ univariate continuous functions~\cite{kolmogorov1957representation}. This is  formalized in the following theorem.

\begin{theo}[Superposition Theorem~\cite{kolmogorov1957representation}]\label{Kolmogorov}
Let $n\in\mathbb{N}$ be a given integer. Any multivariate continuous function $f:\mathbb{R}^n\rightarrow \mathbb{R}$ can represented as a superposition of univariate continuous functions according to:
\begin{align}
\label{multiple_inner_representation}
f(\mathbf{X}) = \sum\limits_{q=0}^{2n} \Phi_q \left(\sum\limits_{p=1}^n \psi_{q,p} (X_p) \right)\ ,   
\end{align}
where $\mathbf{X}\triangleq (X_1,\hdots, X_n)$, $\Phi_q:\mathbb{R}\rightarrow \mathbb{R}$, and $\psi _{q,p}:\mathbb{R}\rightarrow \mathbb{R}$ are univariate continuous functions for $q\in\{0,\dots, 2n\}$ and $p\in\{1,\dots, n\}$. Furthermore, the design of the outer functions $\{\Phi_q\}_q$ depends on $f$, while that of the inner functions $\{\psi_{q,p}\}_{p,q}$ is independent of $f$.
\end{theo}
The decomposition in Theorem~\ref{Kolmogorov} indicates a universal design for the inner functions $\{\psi_{q,p}\}_{p,q}$. The original study in~\cite{kolmogorov1957representation} provided an existence proof for the inner and outer functions. The  results were refined in~\cite{lorentz1962metric} proving that the outer functions $\{\Phi_q\}_q$ can be all replaced by a single function $\Phi$, indicating a significant simplification in designing the outer functions, especially as the data dimension $n$ grows. This result, in turn, was further refined in~\cite{sprecher1965structure} showing that the design of the inner functions can be also simplified by proving that the inner functions $\{\psi_{q,p}\}_{p,q}$ are all the same up to proper scaling and shifts. These results are formalized in the following theorem.

\begin{theo}[Sprecher's Representation \cite{sprecher1965structure}] \label{Sprecher}

Let $n\in\mathbb{N}$ be a given integer. Any multivariate continuous function $f:\mathbb{R}^n\rightarrow \mathbb{R}$ can represented as a superposition of univariate continuous functions according to:
\begin{align}
\label{single_inner_representation}
f(\mathbf{X}) = \sum\limits_{q=0}^{2n} \Phi \left(\sum\limits_{p=1}^n \alpha_p \psi(X_p+qa) +q\right)\ ,  
\end{align}
where $\psi$ is a continuous and monotonically increasing function independent of $f$, and $\Phi:\mathbb{R}\to\mathbb{R}$.
\end{theo}
The representation of $f$ in Theorem~\ref{Sprecher} offers two key implications. First, the decomposition can be uniquely characterized by only two functions $\Phi$ and $\psi$, and $(n+1)$ constants $a,\{\alpha_p\}_p$. Secondly, the design of the inner function $\psi$ is independent of $f$. 
These existence (non-constructive) observations have been more recently complemented with constructive proofs providing algorithms for efficiently determining the inner and outer functions, as well as the attendant constants. In particular, their studies in~\cite{Kurkova91} and~\cite{Kurkova92} provide algorithms that generate an approximate $f$. These algorithms are kernel-based, in which the univariate functions are approximated by linear combinations of the canonical sigmoidal functions. These algorithms are further refined in~\cite{Nakamura} and provide an approximate $f$ that can be made arbitrarily accurate. Subsequent studies in~\cite{Sprecher96,Sprecher97,Koppen2002,braun2009constructive} provide algorithmic and constructive proofs for the KA representation in~\eqref{single_inner_representation}.

\subsection{Distributed Computation}
\label{section:distributed}

\begin{figure}[h!]
    \centering
    \includegraphics[width=0.8\columnwidth]{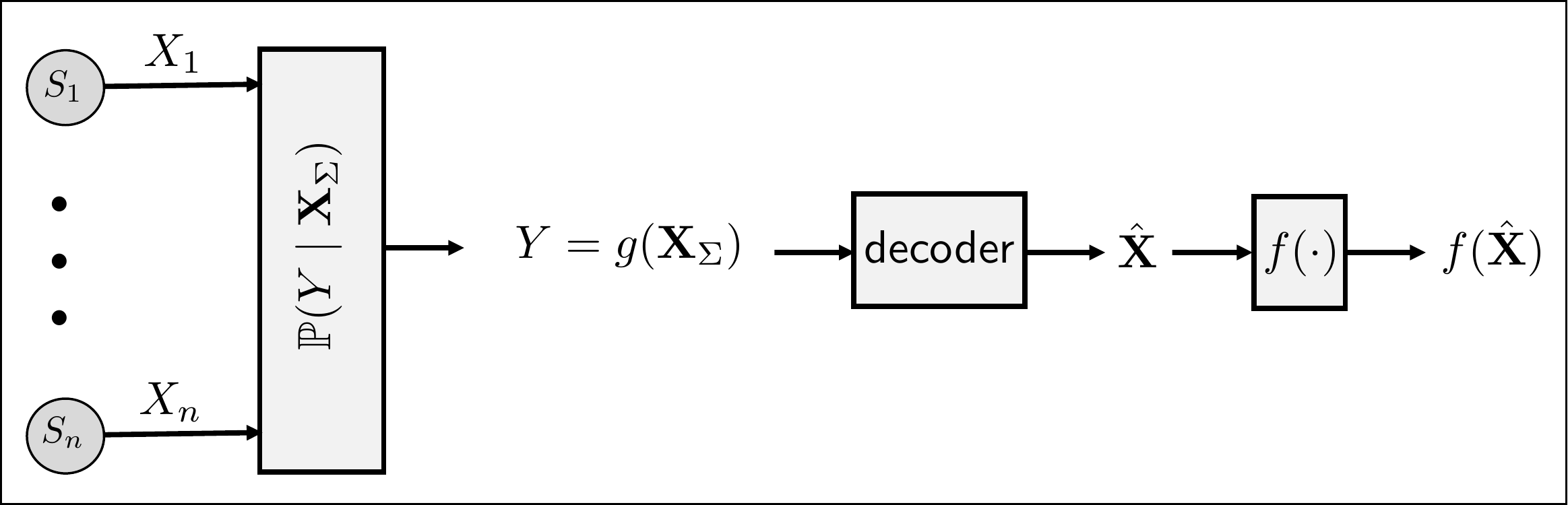}
    \caption{Uncoded transmissions.}
    \label{fig:uncoded}
\end{figure}

The KA decomposition has two key features that make it amenable to distributed computation over wireless channels. Inspired by KA decomposition structures, in this paper, we propose a decomposition-based framework for distributed computing over the additive MAC, e.g., the Gaussian MAC. To furnish context, consider the canonical multiple access channel in which $n$  information sources denoted by $\{S_i\in\{1,\dots,n\}\}$ communicate with a shared receiver. The message of source $S_i$ is denoted by $X_i\in\Omega_i\subseteq\mathbb{R}$. We assume that the messages can be statistically dependent in an arbitrary fashion, and we denote the joint probability measure of $\mathbf{X}$ by $\mathbb{P}_{\mathbf{X}}$. We also define $Y$ as the channel output and denote its probability measure by $\mathbb{P}_{Y}$. By defining 
\begin{align}
\mathbf{X}_\Sigma \triangleq \sum_{i=1}^n X_i\ ,
\end{align}
the output of the additive multiple access channel  is related to the input via a probability kernel denoted by $\mathbb{P}_{Y\;|\; \mathbf{X}_\Sigma}$, rendering
\begin{align}
\mathbb{P}_Y (y)=  \int_{\mathbf{X}\in\prod_{i=1}^n\Omega_i} \mathbb{P}_{Y\;|\; \mathbf{x}_\Sigma}(y\;|\; \mathbf{x})\;  \mathbb{P}_{\mathbf{X}} (\mathbf{x})\; d \mathbf{x}\ ,
\end{align}
where we have defined $\mathbf{X}\triangleq (X_1,\dots, X_n)$. The receiver's objective is to compute $f(\mathbf{X})$, where $f:\mathbb{R}^n\rightarrow\mathbb{R}$ is a continuous function. A direct, albeit inefficient, approach to computing $f(\mathbf{X})$ involves communicating all the raw messages $\mathbf X$ over the MAC; decoding all the individual messages, and generating an estimate $\hat{\mathbf X}$ for $\mathbf X$; and finally computing $f(\hat{\mathbf{X}})$. Clearly, this approach is inefficient since it bears redundancy due to fully communicating $\mathbf X$. We refer to this approach, depicted in Figure~\ref{fig:uncoded}, as  {\em uncompressed transmission}, which serves as a baseline to compare the communication efficiency of the distributed computation frameworks that we design.

In this paper, we propose two distributed computing frameworks that capitalize on decomposing $f(\mathbf X)$ and transmitting the summands contributing to $f(\mathbf X)$ rather than directly transmitting~$\mathbf X$. In the first approach, we show that in the decomposition 
\begin{align}
f(\mathbf{X}) = \sum\limits_{q=0}^{2n} \Phi \left(\sum\limits_{p=1}^n \alpha_p \psi(X_p+qa) +q\right)\ ,  
\end{align}
computing the inner functions $\psi$ can be pushed to the source sites and the outer functions $\Phi$ are computed at the receiver. In this second approach, we provide a graph entropy-based approach to further push computing the outer functions $\Phi$ the source sites. Throughout the rest of the paper, we will focus on the two-user additive MAC for simplicity in describing the ideas and notations. The results can be readily generalized to more than two users with straightforward adjustments. 

\section{Distributed Computing via Compressing the Inner Functions} 
\label{section:inner}

\begin{figure}[h!]
    \centering
    \includegraphics[width=\columnwidth]{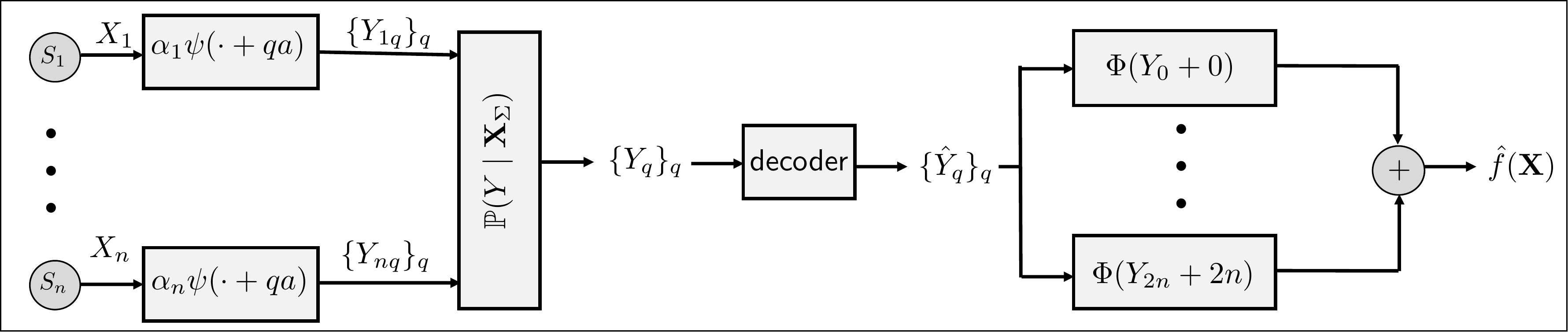}
    \caption{Transmitting inner function representations.}
    \label{fig:innercoded}
\end{figure}

We leverage the decomposition of $f(\mathbf{X})$ in~\eqref{single_inner_representation} to perform part of the computations at the source sites, relieving the sources from communicating the raw messages. This approach translates into reduced communication, imposing less stringent conditions on the rates that the additive MAC should sustain in order to reliably compute $f(\mathbf X)$ at the receiver. Specifically, we decompose the task of computing $f(\mathbf{X})$ to three computing sub-tasks, one carried out by the sources, one by the receiver, and one by the additive nature of the channel. 
\begin{itemize}
\item {\bf Inner functions at the sources:} Computing $f(\mathbf{X})$ involves computing $n$ inner functions of the form $\alpha_p \psi(X_p+qa)$ for $p\in\{1,\dots, n\}$. Computing the term that involves $X_p$ can be entirely carried out by source $S_p$. Hence, computing $\{\alpha_p \psi(X_p+qa): q\in\{0,\dots, 2n\}\}$ is deferred to the source $S_p$, and subsequently, $S_p$ transmits the set $\{\alpha_p \psi(X_p+qa): q\in\{0,\dots, 2n\}\}$ instead of $X_p$.
\item {\bf Superposition by the channel:} The second level of computation involves computing the superposition of the individual terms $\{\alpha_p \psi(X_p+qa): p\in\{1,\dots,n\}\}$ computed by different sources. The additive nature of the channel provides this operation for free, generating the sequence $\{Y_q: q\in\{0,\dots, 2n\}\}$ at the output of the channel, where 
\begin{align}
Y_q\triangleq h\left(U,\; \sum\limits_{p=1}^n Y_{pq}\right)\ , \qquad \mbox{and} \qquad Y_{pq}\triangleq \alpha_p \psi(X_p+qa)\ ,
\end{align}
and $U\in{\cal U}$ is a random variable accounting for a probabilistic mapping from $\{Y_{pq}\}_{p}$ to $Y_q$.
\item {\bf Outer function at the receiver:} Finally, at the receiver the channel output is used to compute $2n+1$ functions $\{\Phi(Y_q+q): q\in\{0,\dots,2n\}\}$. According to~\eqref{single_inner_representation}, the sum of these functions equals to $f(\mathbf{X})$. 
\end{itemize}
The decomposition of $f(\mathbf{X})$ and the subtasks involved are depicted in Figure~\ref{fig:innercoded}. We remark that a key feature of such an approach to distributed computing is that the design of the inner functions $\psi$ is universal and independent of $f$\footnote{A system is called universal with respect to a class of systems if it can compute every function computable by systems in that class (or can simulate each of those systems).}. Hence, the computation sub-task performed by the sources inherits this universality, relieving the sources from changing the inner function $\psi$ as the target function $f$ changes. With changing $f$, Source $S_p$ needs to  update only  the constants  $a$ and $\alpha_p$. This is especially an imperative feature when the target function $f$ to be computed changes frequently. 
In practice, link losses and interference restrict the flexibility of KA decomposition. We seek to devise new achievable coding principles for distributed computation of nonlinear functions in networks, to use KA decomposition more flexibly in topologically-constrained settings. To do so, we will explore whether KA decomposition can provide a universal encoding mechanism via the uniqueness of inner functions.  

In the remainder of this section, we provide three examples of distributed computing, where the KA representation of the function of interest takes a closed-form for the inner and out functions $\psi$ and $\Phi$. In this section, the emphasis is placed on incorporating the inner functions into the compression strategy by moving them to the source sites. The outer functions will be implemented at the receiver.

\subsection{Computing Multiplication}
\label{section:multiplication_inner}
We first consider the problem of computing the absolute value of the product of the inputs at the receiver, i.e., 
\begin{align}
    f(\mathbf{X}) = \left|\prod_{i=1}^n\; X_i\right| \ .
\end{align}
Following the canonical KA representation, $f(\mathbf{X})$ can be equivalently represented as
\begin{align}
  f(\mathbf{X})= \sum\limits_{q=0}^{2n} \Phi \left(\sum\limits_{p=1}^n \alpha_p \psi(X_p+qa) +q\right)\ ,
\end{align}
by setting
\begin{align}
    a=0\ , \quad \alpha_p=1\ , \;\forall p\in\{1,\dots, n\}\ , \quad \psi(x)=\log|x|\ , \quad \Phi(x)=\frac{e-1}{e^{2n+1}-1}\exp(x)\ .
\end{align}
It is noteworthy that based on this decomposition, the information that source $S_p$ should transmit, i.e., $\{Y_{pq}\}_q$, where $Y_{pq}= \alpha_p\psi(X_p+qa)$, contains $2n+1$ repetitions of $\psi(X_p)=\log|X_p|$. Operationally, each source can avoid such repetitions and instead transmits only one of these elements. This leads to the following simplified representation of $f(\mathbf X)$
\begin{align}
    f(\mathbf X)=\exp\left(\sum_{i=1}^n\log|X_i| \right)\ , 
\end{align}
fitting the Kolmogorov's original representative in~\eqref{multiple_inner_representation} by setting for all $p\in\{1,\dots, n\}$ and $q\in\{0,\dots, 2n\}$ according to:
\begin{align}
    \psi_{pq}(x)=\mathb{1}_{\{q=0\}}\cdot  \log|x|\ , \quad \mbox{and} \quad  \Phi_q(x)=\mathb{1}_{\{q=0\}}\cdot\exp(x)\ .
\end{align}
It can be readily verified that the rates required for communicating $\{Y_{pq}\}_{pq}$ is less stringent than those required by directly transmitting $\{X_p\}$  with Slepian-Wolf compression. Specifically, note that for any subset of sources ${\cal S}\subseteq\{1,\dots, n\}$, for direct communication, the communication rates must satisfy
\begin{align}\label{eq:sum1}
\sum\limits_{p\in\mathcal{S}}R_p\geq H\left(\{X_p: p\in\mathcal{S}\} \; \vert \;  \{X_p: p\notin\mathcal{S}\}\right) =  H\left(\{X_p: p\in\mathcal{S}\} \right)\ .
\end{align}
On the other hand, when transmissions are designed to be compressed based on the structures of the inner functions, the communication rates must satisfy
\begin{align}\label{eq:sum2}
\sum\limits_{p\in\mathcal{S}}R_p\geq H\left(\{Y_{pq}: p\in\mathcal{S}, q\in\{0,\dots, 2n\}\}\right) = H\left(\{Y_{p0}: p\in\mathcal{S}\}\right)\ .
\end{align}
Noting that
\begin{align}
    H\left(\{Y_{p0}: p\in\mathcal{S}\}\right) = H\left(\{\log|X_p|: p\in\mathcal{S}\}\right) \leq H\left(\{X_p: p\in\mathcal{S}\}\right)\ ,
\end{align}
indicates that the required region of rates in~\eqref{eq:sum2} is confined in the region specified by~\eqref{eq:sum1}. As an example, consider that the $n$ source messages $\{X_p:\,p\in\{1,\dots, n\}\}$ are independent and identically distributed according to a uniform distribution over the shared alphabet 
\begin{align}
    {\cal X}\triangleq \{-M,\dots, M\}\ ,
\end{align}
for an integer $M\in\mathbb{N}$. Hence, for any subset of sources ${\cal S}\subseteq\{1,\dots, n\}$ we have
\begin{align}
\label{entropy_source_multiplication}
    H\left(\{X_p: p\in\mathcal{S}\}\right) & = |{\cal S}|\log(2M+1)\ , \\
\label{entropy_innerfunction_multiplication}    
  \mbox{and} \qquad H\left(\{Y_{p0}: p\in\mathcal{S}\}\right) & = |{\cal S}|\left(\log(2M+1)-\frac{2M}{2M+1}\right)\ ,
\end{align}
indicating compressing by accounting for the inner functions necessitates strictly lower communication rates. This gain is due to (i) incorporating the structure of $f(\mathbf X)$ for designing transmitted signals; (ii) and leveraging the additive structure of the channel.



\subsection{Computing the \texorpdfstring{$\ell_m$}{ellm}-norm}
\label{section:polynomial_simpler}
In the second example, we examine computing the $\ell_m$-norm of $\mathbf X$ at the receiver, i.e., \begin{align}
    f(\mathbf X) = \left(\sum_{i=1}^n |X_i|^m\right)^\frac{1}{m}\ .
\end{align}
This function be represented based on Kolmogorov's original representation by setting for all $p\in\{1,\dots, n\}$ and $q\in\{0,\dots, 2n\}$ according to:
\begin{align}\label{eq:power}
    \psi_{pq}(x)=\mathb{1}_{\{q=0\}}\cdot  |x|^m\ , \quad \mbox{and} \quad  \Phi_q(x)=\mathb{1}_{\{q=0\}}\cdot x^\frac{1}{m}\ .
\end{align}
Similarly to the case of computing the product in Section~\ref{section:multiplication_inner}, the region of communication rates for computing $f(\mathbf X)$ by using the decomposition in~\eqref{eq:power} and computing the inner functions at the source sites is confined within the region of rates necessary for communicating the sources.


\subsection{Computing Polynomials}
\label{section:polynomial_inner}

As the last example, we provide an example in which KA decomposition has no gains over direct communication of the messages. This is intended to showcase that gains in communication systems are not always guaranteed, even though in general for most functions they are viable. This observation will also serve as a reference point to show that for the same function, compressing the sources based on the structures of both inner and outer functions does provide gains over direct communication. We will discuss this in Section~\ref{section:innerandouter}. For this purpose, we consider computing the polynomial functions
\begin{align}
    f(\mathbf X) = \left(\sum_{i=1}^n X_i\right)^m\ ,
\end{align}
which can be represented based on Kolmogorov's original representation by setting for all $p\in\{1,\dots, n\}$ and $q\in\{0,\dots, 2n\}$ according to:
\begin{align}\label{eq:power2}
    \psi_{pq}(x)=\mathb{1}_{\{q=0\}}\cdot  x\ , \quad \mbox{and} \quad  \Phi_q(x)=\mathb{1}_{\{q=0\}}\cdot x^m\ .
\end{align}
It can be readily  verified that for any subset of sources ${\cal S}\subseteq\{1,\dots, n\}$:
\begin{align}
    H\left(\{Y_{p0}: p\in\mathcal{S}\}\right) = H\left(\{X_p: p\in\mathcal{S}\}\right)\ .
\end{align}
This indicates that no gain in applying KA decomposition in the form adopted in~\eqref{eq:power2}. We will revisit this setting again in Section~\ref{section:power2}.



\section{Distributed Computing via Compressing the Inner \& Outer Functions}
\label{section:innerandouter}

The framework in Section~\ref{section:inner} only partially leverages the superposition representation of $f(\mathbf X)$. Specifically, it does not account for the outer functions for compressing the messages and generating the channel inputs. This indicates that the full potential of the decomposition for distributed compression is not properly exploited. In this section, we provide an alternative distributed compression framework that also accounts for the outer functions at the source sites. Bringing the outer functions to the source sites, however, is not directly viable. Specifically, this will require that each source has non-causal access to the messages of all other sources to be able to compute the terms $\Phi(Y_q+q)$. To circumvent this, we will deploy an approach based on graph entropy to capture the correlation between the information sources, and use that as a surrogate accounting for the computing tasks expected from the outer functions. We start by providing an overview of the notion of graph entropy as a means for effectively capturing the statistical dependence of the sources, based on which we will later describe the framework.

\subsection{Graph Entropy: Overview}
\label{section:graph}
\begin{figure}[h!]
    \centering
    \includegraphics[width=0.9\columnwidth]{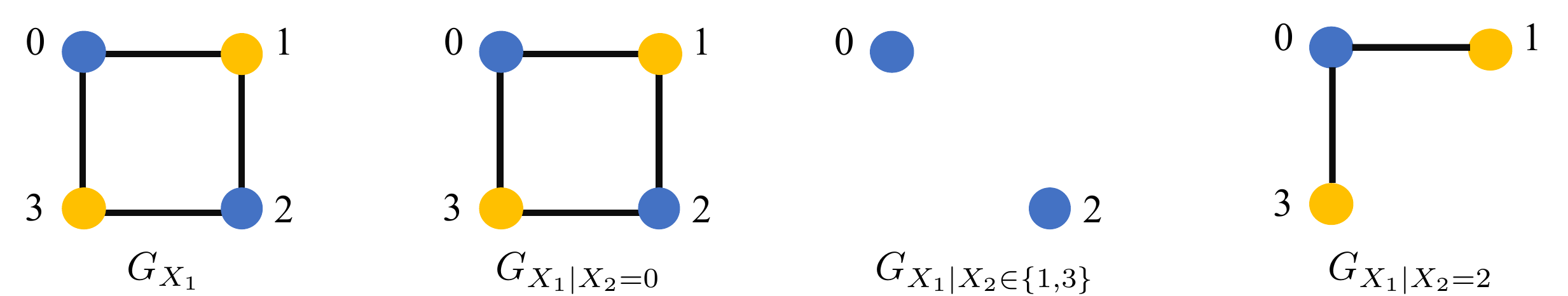}
    \caption{A characteristic graph $G_{X_1}$ that source 1 with $X_1\in\{0,1,2,3\}$ builds to compute $f(\mathbf{X})=\underset{p\in \{1,\dots,n\}}{\xor} X_p$. Conditional characteristic graphs for the same source given $X_2$.}
    \label{fig:characteristicgraph}
\end{figure}

In functional compression of source variables $\mathbf{X}$ to recover $f(\mathbf{X})$ at the receiver, each source encodes its output by creating a collection of equivalence classes such that each class generates a distinct outcome of $f$, independent of the possible realizations of the other source random variables. More precisely, each source has the knowledge of the function, and it assigns different codes to the values of source random variables that need to be distinguishable at the receiver. To determine which pairs of the random variable realizations should be assigned to different codes,  \cite{korner1973coding} and \cite{feizi2014network} devise a graphical model. According to this model, corresponding to each random variable (source) one graph is generated, which is referred to as the \emph{characteristic graph} of that random variable~\cite{shannon1956zero} and \cite{korner1973coding}. We denote the graph associated with $X_p$ by $G_{X_p}$, for $p\in\{1,\dots,n\}$. The vertices of graph $G_{X_p}$ represent the distinct possible realizations of $X_p$. There will be an edge between two vertices if they should be distinguished at the receiver. In the context of computing $f(\mathbf X)$, in graph $G_{X_1}$, the vertices of two different realizations of $X_1$ are connected if these two realizations lead to different function value $f(x_1,X_2,\dots, X_n)$ for all other realizations of  $\{X_p:p\in\{2,\dots, n\}\}$ to different function values. We illustrate a characteristic graph example in Figure \ref{fig:characteristicgraph}. In this example, the receiver aims to recover 
\begin{align}
f(\mathbf{X})=\underset{p\in \{1,\dots,n\}}{\xor} X_p\,,    
\end{align}
where $\xor$ denotes the binary exclusive or operation. The vertices of $G_{X_1}$ are the realizations of $X_1 \in \{0,1,2,3\}$. There do not exist edges between vertices $0$ and $2$, and between $1$ and $3$ because they lead to the same function outcomes. Hence, two distinct colors suffice to describe $G_{X_1}$. We also illustrate the conditional characteristic graph $G_{X_1|X_2=x_2}$ for $x_2\in\{0,1,2,3\}$, where given $X_2=0$, the support set of $X_1$ is unaffected. However, if $X_2$ is odd-valued, then $X_1$ is even-valued, i.e., conditional on $X_2\in\{1,3\}$, for $X$ we have $X_1\vert X_2\in\{0,2\}$. Furthermore, $X_1\vert \{X_2=2\}\in\{0,1,3\}$~\cite{Sefidgaran:ISIT2011}.

\begin{figure}[h!]
    \centering
    \includegraphics[width=.5\columnwidth]{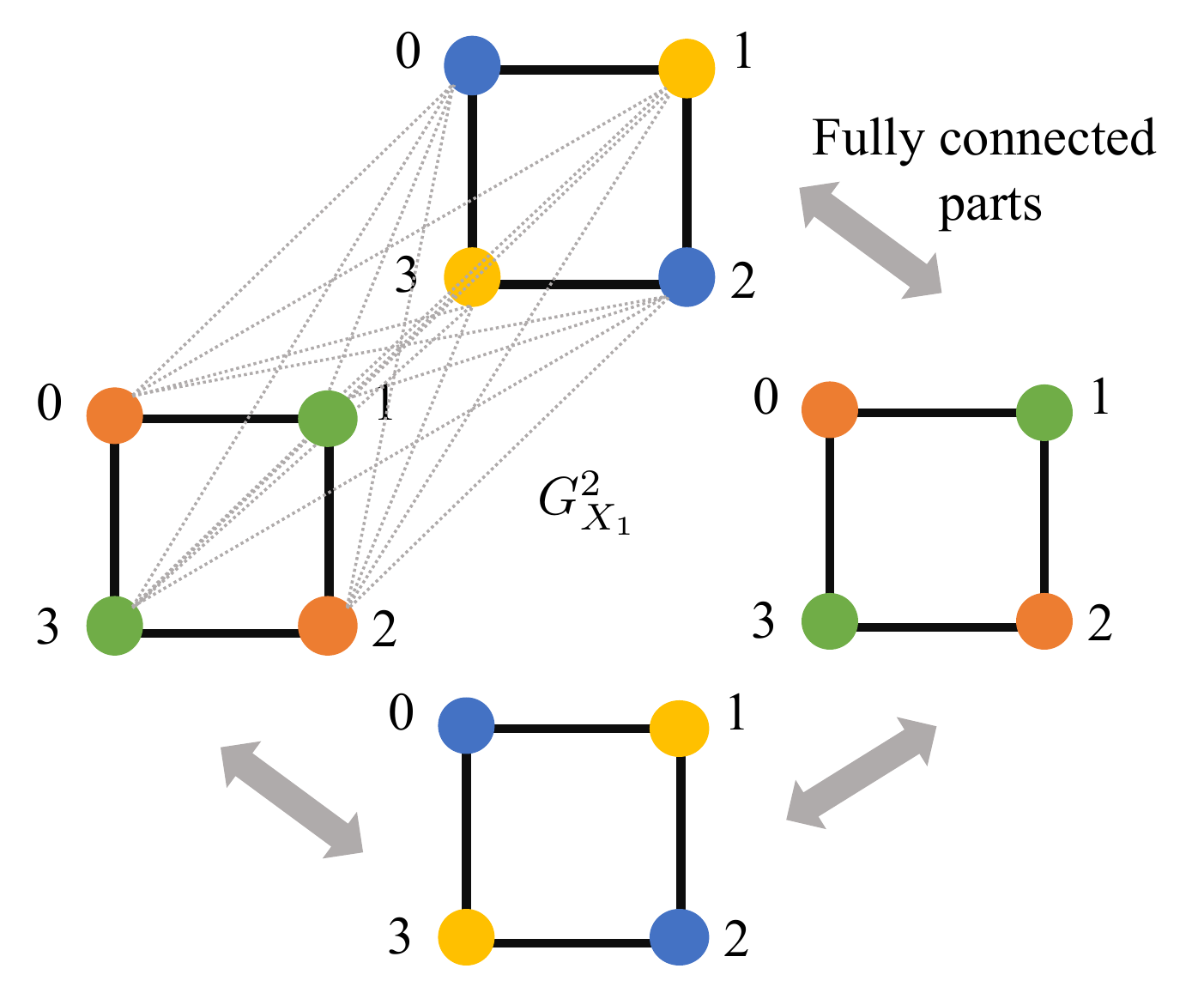}
    \caption{Power graph $G^2_{X_1}$ that source $S_1$ with $X_1^2\in\{0,1,2,3\}^2$ builds to compute $f(\mathbf{X})=\underset{p\in \{1,\dots,n\}}{\xor} X_p$.}
    \label{fig:powergraph}
\end{figure}

Corresponding to the characteristic graph $G_{X_p}$, we define the graph entropy of source $S_p$, denoted by $H_{G_{X_p}}(X_p)$, as~\cite{feizi2014network}:
\begin{align}
\label{GraphEntropy}
H_{G_{X_p}}(X_p)\triangleq   \min_{X_p\in W_p\in \Gamma(G_{X_p})} I(X_p,W_p) \ ,
\end{align}
where $\Gamma(G_{X_p})$ is the set of all maximal {\em independent sets} of $G_{X_p}$. It is noteworthy that the independent set of a graph is a subset of its vertices no two of which are adjacent. Accordingly, a maximal independent set is an independent set that is not a proper subset of any other independent set. Accordingly, we define $H_{\{G_{X_p}\}_{p\in{\cal S}}}(X_{\cal S})$ as the joint graph entropy of sources in set~${\cal S}$. For example for ${\cal S}=\{1,2\}$ we have $H_{\{G_{X_p}\}_{p\in{\cal S}}}(X_{\cal S})=H_{G_{X_1},G_{X_2}}(X_1,X_2)$. This notion of graph entropy can be extended to graph conditional entropy. Specifically, corresponding to sources $S_p$ and $S_r$ we define the graph conditional entropy as
\begin{align}
   H_{G_{X_p}} (X_p\vert X_r )\triangleq\min_{\overset{X_p\in W_p\in \Gamma(G_{X_p})}{ W_p-X_p-X_r}} I(W_p;X_p\vert X_r)\ ,
\end{align}
where $G_{X_p}$ is the conditional characteristic graph that $X_p$ builds given the knowledge of $X_r$. 
Based on these definitions, for any subset of sources ${\cal S}\subseteq\{1,\dots, n\}$, we define $X_{\cal S}\triangleq \{X_p:\, p \in {\cal S}\}$. By denoting the complement of ${\cal S}$ by ${{\cal S}^c}$, the rate region for the functional compression problem is lower bounded according to~\cite{feizi2014network}:
\begin{align}
\label{functional_compression_rate}
\sum\limits_{p\in {\cal S}}R_p\geq  H_{\{G_{X_p}\}_{p\in{\cal S}}} (X_{\cal S}| X_{{\cal S}^c})\ .
\end{align}
For a given characteristic graph $G_{X_p}$ we assign labels, also known as colors, to the vertices of $G_{X_p}$ to distinguish the possible function outcomes of variable $X_p$ for any given set of variables $\{X_r,\, r\neq p\}$. We consider a vertex coloring scheme such that no two adjacent vertices share the same color. In other words, adjacent vertices yield different values of the same function.    
For a given graph $G_{X_p}$, we specify a valid coloring of the graph as follows. Let $C(X_p)$ be a random variable that achieves a valid coloring of the characteristic graph $G_{X_p}$. We define \emph{chromatic entropy} as the minimum entropy of $G_{X_p}$ over all possible valid coloring schemes~\cite{alon1996source}. The chromatic entropy is defined as
\begin{align}
\label{chromatic}
H^{\chi}_{G_{X_p}}(X_p)\triangleq \min\Big\{H(C(X_p)):\, C(X_p) \mbox{ is a coloring of } G_{X_p} \Big\}\,.   
\end{align}
The graph entropy and the chromatic entropy are related according to~\cite{feizi2014network}:
\begin{align}
\label{chromatic_vs_characteristic}
H_{G_{X_p}}(X_p)=\lim\limits_{n\to \infty}\frac{1}{n}H^{\chi}_{G^n_{{\bf X}_p}}({\bf X}_p)\,,
\end{align}
where $G^n_{{\bf X}_p}$ denotes the $n^{\rm th}$ power of a graph $G_{X_p}$ where its vertices are the elements of $\mathcal{X}^n$ and there exists an edge between any pair of vertices provided that there exists at least one coordinate $i$ of vectors ${\bf x}_1,{\bf x}_2 \in \mathcal{X}^n$ such that $(x_{1i},x_{2i})$ is in the edge set of $G_{X_p}$. We denote a valid coloring of $G^n_{{\bf X}_p}$ by $C({\bf X}_p)$. We note that (\ref{chromatic_vs_characteristic}) implies that for computing a function of discrete memoryless sources of data at the receiver, a source can apply a modularized encoding approach that consists of the following two processes: (i)  source $S_p$ finds a coloring for $G^n_{{\bf X}_p}$, and (ii) it encodes the colors using the Slepian-Wolf encoding scheme, which achieves the entropy bound of the coloring variable $C({\bf X}_p)$. In Figure \ref{fig:powergraph} we illustrate the $2^{\rm nd}$ power graph of $G_{X_1}$ for the example in Figure \ref{fig:characteristicgraph}. Note that the four graphs indicated with solid colors are fully connected to each other. This scheme requires four distinct colors to describe $G^2_{{\bf X}_1}$. For further details on graph entropies and coloring of characteristic graphs, we refer the reader to \cite{korner1973coding,alon1996source,orlitsky2006coding}.

\begin{figure}[h!]
    \centering
    \includegraphics[width=\columnwidth]{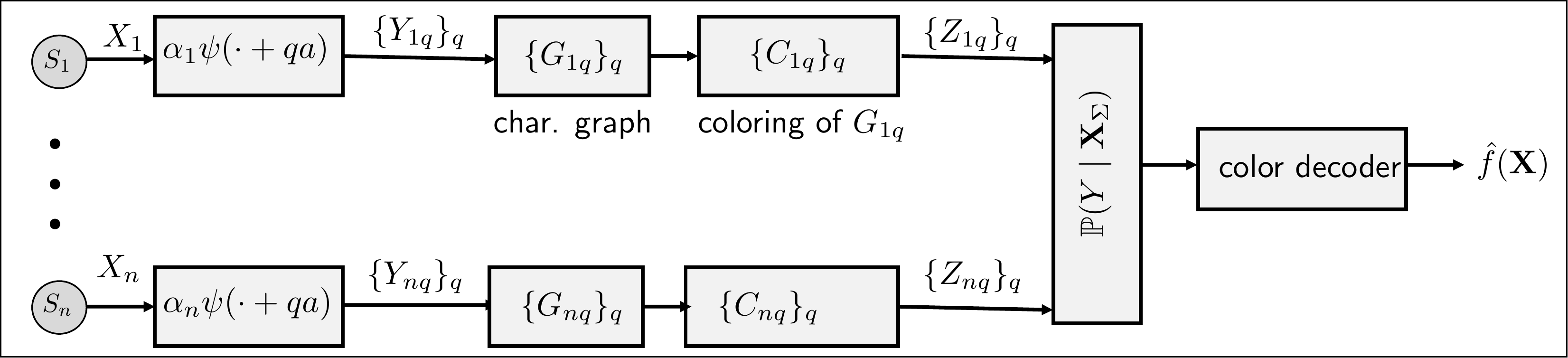}
    \caption{Joint functional compression.}
    \label{fig:jointcoded}
\end{figure}
In Figure \ref{fig:jointcoded} we illustrate the modularized graph coloring approach for joint compression for computing a class of KA functions. For ease of exposition, we dropped the superscripts from the colorings and the characteristic graphs. In this scheme, source $p\in \{1,\dots,n\}$ builds $2n+1$ characteristic graphs, where $G_{pq}$, $q\in\{0,\dots,2n\}$ has a set of vertices that are possible values of $X_p$. 
Following the KA decomposition of $f(\mathbf X)$ in~\eqref{multiple_inner_representation}, we define $G_{pq}$ as the characteristic graph that source $p$ builds for recovering the function 
\begin{align}
   \Phi \left(\sum\limits_{p=1}^n \alpha_p \psi(X_p+qa) +q\right)\ ,
\end{align}
and denote the associated coloring by $Z_{pq}$. It is noteworthy that while such construction of the graphs provides a systematic mapping between KA decomposition and building the graphical models and their coloring, it can be modified to become more efficient. Specifically, in the modified approach, each source $S_p$ generates only one characteristic graph denoted by $G_p$ with coloring $Z_p$. In this approach, besides having a fewer number of graphs (one per source), each graph $G_p$ has a fewer number of edges compared with $G_{pq}$. A reduced number of edges translates to significant savings in compression over the scheme based on directly implementing the KA decomposition.

In the remainder of this section, we provide several examples in the context of which we provide more details on the design of the compression schemes. Furthermore, we compare the compression rates achieved based on this method and those achieved by compressing only the inner functions (Section~\ref{section:inner}) as well as direct communication of the sources. For simplicity in notations and describing graph generating processes, we focus on the two-user setting (i.e., $n=2$) decomposition with one outer function, i.e., $f(X_1,X_2)=\Phi(\psi(X_1)+\psi(X_2))$. We will consider two distinct settings for $\Phi$, where in one set of settings $\Phi$ is additively separable and in the other set it is not. When $\Phi$ is an additively separable functions, we can write $f(X_1,X_2)=\Phi(Y_1+Y_2)=g_1(Y_1)+g_2(Y_2)$. 

\subsection{Computing Multiplication}
\label{section:multiplication}
As for the first example, we revisit the multiplication function, i.e.,
\begin{align}
    f(\mathbf X)=\left|\prod_{i=1}^n\; X_i\right|\ .
\end{align}
In the simplified KA decomposition form, this can be represented as
\begin{align}
    f(\mathbf X)=\Phi\left(\sum_{i=1}^n\psi(X_i)\right)\ , \quad \mbox{where} \quad \Phi(x)=2^x \quad \mbox{and} \quad \psi(x)=\log x\ .
\end{align}

To start, we first note that unlike computing the absolute value of the product in~Section~\ref{section:inner}, computing the product by only compressing the inner functions does not have any communication gain over the direct transmission of $|X_p|$ for all $p\in\{1,\dots,n\}$. This can be readily verified by following the same line of argument as in Section~\ref{section:multiplication_inner}. In this subsection, however, we show that a compression that also accounts for the outer functions does lead to strictly improving the communication rate requirements. For this purpose, we assume that sources are uniformly and identically distributed over ${\cal X}= \{-M,\dots, M\}$. The entropies of the collection of variables $\{X_{p}: p\in\mathcal{S}\}$ and $\{Y_{p}: p\in\mathcal{S}\}$ satisfy (\ref{entropy_source_multiplication}) and (\ref{entropy_innerfunction_multiplication}), respectively, given in~Section~\ref{section:inner}. We note that for all $p\in\{1,\dots,n\}$
\begin{align}
|X_p|\in\{0,1,\dots,M\}\,, \quad p_{|X_p|}=\frac{1}{2M+1}\times(1,2,\dots,2)\,.    
\end{align}
Note that there is a bijection between $|X_p|$ and $Y_{p}$ and we use them interchangeably. This indicates that $Z\triangleq f(\mathbf X)$ is a product distribution with a probability mass function (pmf) 
\begin{align}
p_{Z}(z)=\begin{cases}
1-\left(\frac{2M}{2M+1}\right)^n, &   z= 0\ ,\nonumber\\
\left(\frac{2}{2M+1}\right)^n \sum\limits_{\{|x_p| : p\in\mathcal{S}\}} \mathbbm{1}{\left(\prod\limits_{p=1}^n |x_p|=z\right)},& z\neq 0 ,\quad p=1,\dots,n\ \nonumber
\end{cases}\ .
\end{align}

\begin{figure}[h!]
    \centering
    \includegraphics[width=\columnwidth]{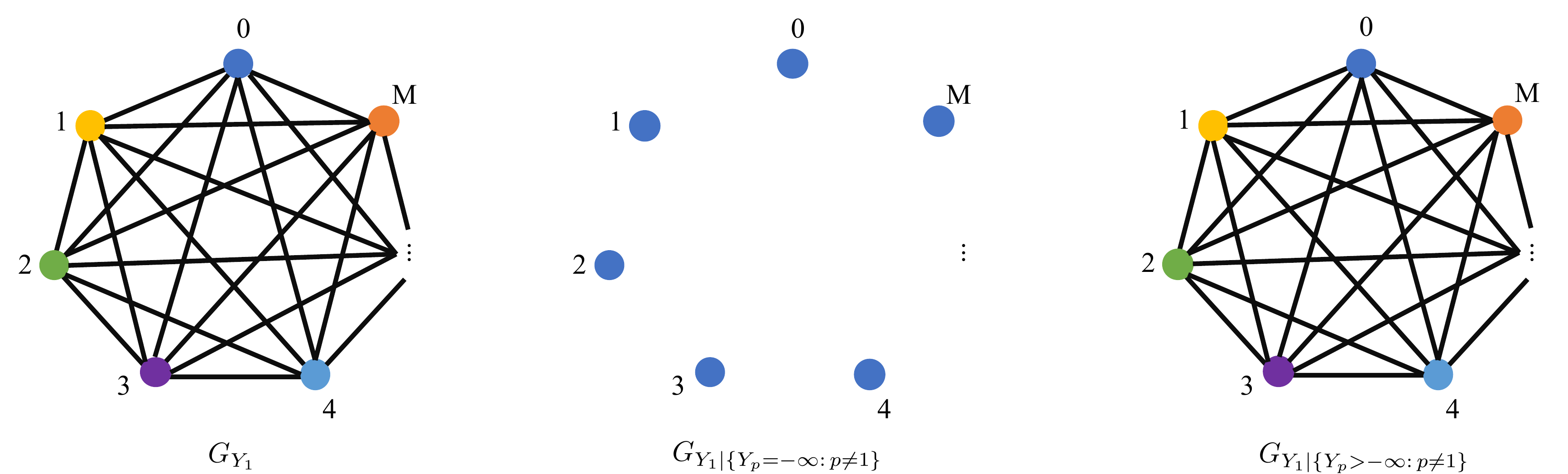}
    \caption{Characteristic graphs for multiplication.}
    \label{fig:product_graph_example}
\end{figure}

Since all the outcomes of $Y_p$ need to be distinguished at the source, the characteristic graphs satisfy $H_{G_{Y_p}}(Y_p)=H(Y_p)$. It is easy to note that  $H_{\{G_{Y_p}\}_{p\in\mathcal{S}}}(Y_{\mathcal{S}})<H(\{Y_p:p\in\mathcal{S}\})=|\mathcal{S}| H(Y_p)$. For any subset of sources $p\in\mathcal{S}$, the conditional graph entropies are given by:
\begin{align}
&H_{\{G_{Y_p}\}_{p\in\mathcal{S}}}\Big(\{Y_p: p\in\mathcal{S}\}\; \vert \;  \{Y_p: p\notin\mathcal{S}\}\Big)\nonumber\\
&=\mathbb{P}\left(\prod\limits_{p\notin\mathcal{S}} |X_p|\neq 0\right) H_{\{G_{Y_p}\}_{p\in\mathcal{S}}}\Big(\{Y_p: p\in\mathcal{S},\, Y_p>-\infty\}\Big)\nonumber\\
&\leq\left(\frac{2M}{2M+1}\right)^{n-|\mathcal{S}|}|\mathcal{S}| H_{G_{Y_p}}(Y_p)\,, \nonumber    
\end{align}
where the inequality is because $\{G_{Y_p}\}_{p\in\mathcal{S}}$ are jointly distributed. Hence, the graph entropy approach provides an additional communication gain over the approach in Section~\ref{section:inner}. The characteristic graphs are shown in Figure \ref{fig:product_graph_example}.

Note that unlike computing the absolute value of the product in~Section~\ref{section:inner}, computing the product by only compressing the inner functions does not have any communication gain over the direct transmission of $\{X_i:i\in\{1,\dots,n\}\}$. For instance, consider the simple setting of $n=3$. By following the same line of arguments as in Section~\ref{section:multiplication_inner}, it can be readily verified that computing the product by only compressing the inner functions does not have any communication gain over the direct transmission of $\{X_1,X_2\}$.

\begin{figure}[h!]
    \centering
    \includegraphics[width=0.9\columnwidth]{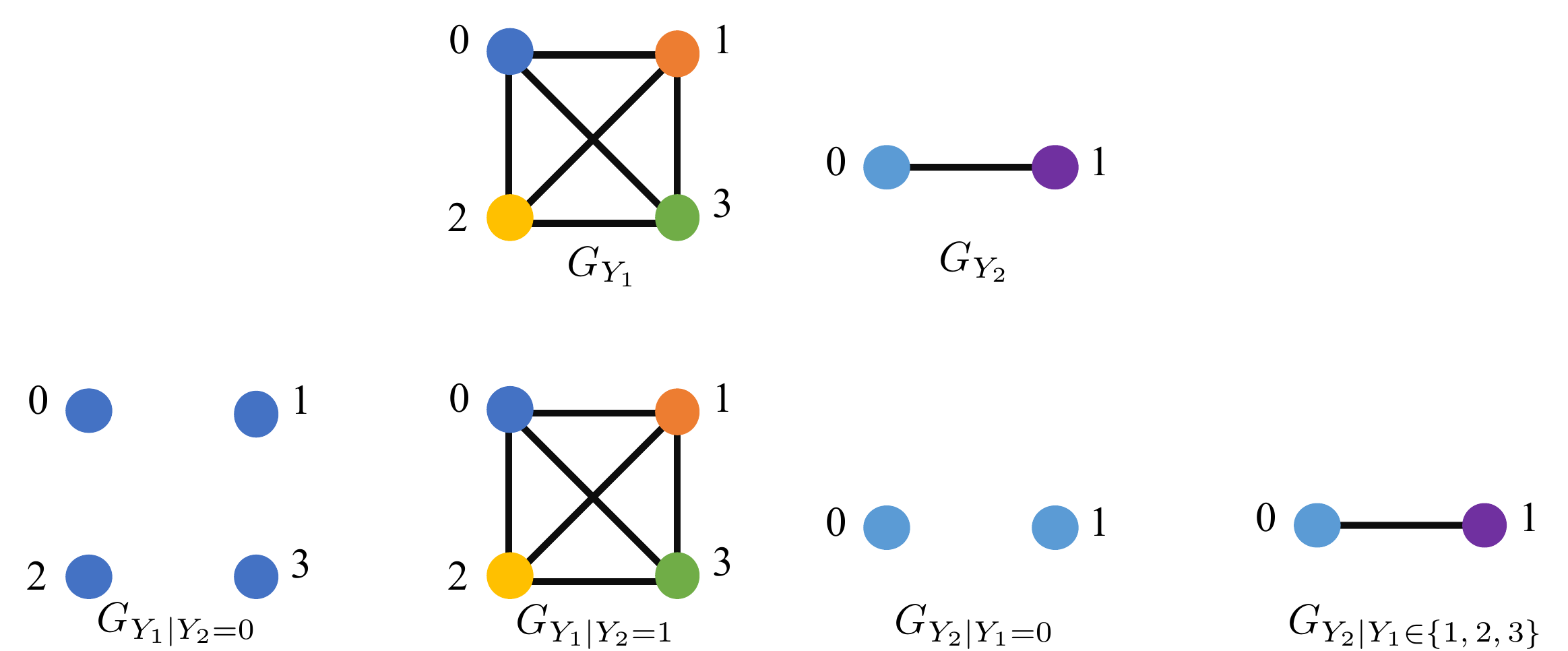}
    \caption{Characteristic graphs for distributed multiplication.}
    \label{fig:product_example}
\end{figure}

In this subsection, however, we show that a compression that also accounts of the outer  functions does lead to strictly improving the communication rate requirements. For this purpose, we consider the following alphabets for $X_1$ and $X_2$:
\begin{align}
    X_1\in\{0,1,2,3\}\ , \quad \mbox{and} \quad X_2\in \{0,1\}\ .
\end{align}
$X_1$ and $X_2$ are assumed to be independent and have uniform distributions. This indicates that $f(\mathbf X)\in \{0,1,2,3\}$ with a probability mass function (pmf) at the respective coordinates $\frac{1}{8}\times(5,1,1,1)$. Hence, 
\begin{align}\label{eq:product_outer1}
H(X_1)=2\ , \quad H(X_2)=1\ , \quad H(X_1,X_2)=3\ ,\quad \mbox{and} \quad H(f(X_1,X_2))=1.5488 \ .   
\end{align}
Next, we provide the characteristic graphs associated with the graphs for the multiplication function in Figure~\ref{fig:product_example}. A coloring-based approach to compressing $\mathbf X$ requires the following set of rates 
\begin{align}\label{eq:product_outer2}
H_{G_{X_1}}(X_1\vert X_2)=1 \ , \quad H_{G_{X_2}}(X_2\vert X_1)=\frac{3}{4}\,, \quad\mbox{and} \quad H_{G_{X_1},G_{X_2}}(X_1,X_2)=2\ .
\end{align}
Comparing \eqref{eq:product_outer1} and \eqref{eq:product_outer2} shows that an uncoded transmission scheme and the scheme that only compresses the outer functions (Section \ref{section:inner}) require 
\begin{align}
    R_1\geq 2\,, \quad R_2\geq 1\,, \quad \mbox{and} \quad R_1+R_2\geq 3\ ,
\end{align}
whereas compressing both inner and outer functions based on graph entropies requires
\begin{align}
    R_1\geq 1\,, \quad R_2\geq \frac{3}{4}\,, \quad \mbox{and} \quad R_1+R_2\geq 2\ .
\end{align}
It is also noteworthy that every color pair received is a valid function outcome, but a unique function outcome might have multiple color pairs due to the distributed setting. This leads to $H_{G_{X_1},G_{X_2}}(X_1,X_2)>H(f(\mathbf{X}))$.

\subsection{Computing the \texorpdfstring{$\ell_m$}{ellm2}-norm}
\label{section:norm}
As our second example, we consider the $\ell_m$-norm function 
\begin{align}
f(\mathbf X)=\left(\sum_{i=1}^n |X_i|^m\right)^{\frac{1}{m}}\ .   
\end{align}
In the simplified KA decomposition form, this can be represented as
\begin{align}
    f(\mathbf X)=\Phi\left(\sum_{i=1}^n \psi(X_i)\right)\ , \quad \mbox{where} \quad \Phi(x)=x^{\frac{1}{m}} \quad \mbox{and} \quad \psi(x)=|x|^m\ .
\end{align}
To demonstrate the savings of different compression schemes and how they are affected by the source distribution, we consider two special cases.

\begin{enumerate}[(i)]
\item {\bf Independent sources.} The sources are independent and uniformly distributed over the alphabet $\mathcal{X}=\{-M,\dots, M\}$. Hence, the inner transformations of the variables denoted by $Y_p=\psi(X_p)$ for $p\in\{1,\hdots,n\}$, take values in $\{0,1,2^m,\dots,M^m\}$ with the pmf $\frac{1}{2M+1}\times(1,2,\dots,2)$. For this setting, for any subset of sources $\mathcal{S}\subseteq \{1,\dots, n\}$ we have
\begin{align}
\label{source_entropy_l2_norm}
H(\{X_p:p\in\mathcal{S}\})&=|\mathcal{S}|\log(2M+1)\ , \\
  \mbox{and} \qquad H(\{Y_p:p\in\mathcal{S}\})&=|{\cal S}|\left(\log(2M+1)-\frac{2M}{2M+1}\right)\,.
\end{align}
We note that the characteristic graphs satisfy $H_{G_{Y_p}}(Y_p)=H(Y_p)$ and
\begin{align}
H_{\{G_{Y_p}\}_{p\in\mathcal{S}}}(\{Y_p: p\in\mathcal{S}\}\; \vert \;  \{Y_p: p\notin\mathcal{S}\})=|\mathcal{S}| H_{G_{Y_p}}(Y_p) \, ,\nonumber    
\end{align}
due to the independence among sources, where we note that  $H_{\{G_{Y_p}\}_{p\in\mathcal{S}}}(Y_{\mathcal{S}})=H(\{Y_p:p\in\mathcal{S}\})=|\mathcal{S}| H(Y_p)$. Hence, the graph entropy approach does not provide an additional gain over the approach in Section~\ref{section:inner}.

\begin{figure}[h!]
    \centering
    \includegraphics[width=0.45\columnwidth]{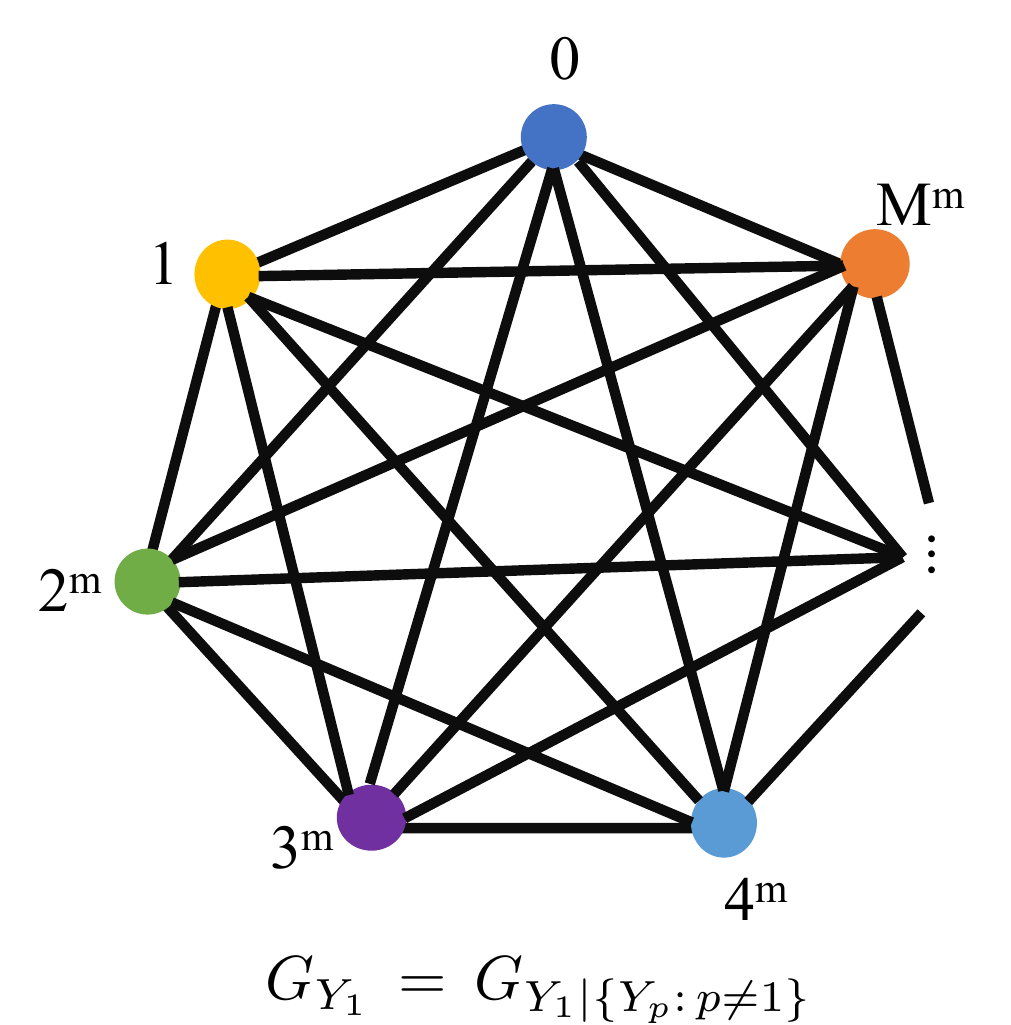}
    \caption{Characteristic graph for $\ell_m$-norm.}
    \label{fig:m_norm_example}
\end{figure}

\item {\bf Correlated sources.} We let $\psi(X_1)\in \{0,1,2^m\}$ and $\psi(X_2)\in \{0,1,2^m\}$ have the following ordered marginal pmfs $\frac{1}{3}(1,1,1)$ and $\frac{1}{4} (2,1,1)$, respectively. We also let $Y_1\sim \psi(X_1)$ and $Y_2\sim \psi(X_2)$ have the same marginal pmfs as $X_1$ and $X_2$, respectively. 
We further assume that $\mathbb{P}(Y_1=Y_2)$ is maximized, which  ensures that $H(Y_1,Y_2)\approx H(Y_1)$ and given that $Y_1$ is sent, the marginal rate needed for $Y_2|Y_1$ is small. Maximizing $\mathbb{P}(Y_1=Y_2)$ can be achieved via maximal coupling. In other words, we can write these variables as
\begin{align}
    Y_1&=UT+(1-U)V\ , \\
    Y_2&=UT+(1-U)W\ ,
\end{align}
where $U\sim\mbox{Bern}(1-\delta)$ and $\delta=\frac{1}{6}$, and the discrete random variables $T,\,V$, and $W$ have  pmfs 
\begin{align}
p_T=\frac{1}{10}(4,3,3)\ , \quad p_V=\frac{1}{2}(0,1,1)\,    , \quad \mbox{and} \quad p_W=(1,0,0)\ .
\end{align}
Hence, their entropies are $H(T)=1.571$, $H(V)=1$, and $H(W)=0$. In this setting, the joint entropy of the variables is
\begin{align}
\label{joint_entropy_l2_norm}
    H(Y_1,Y_2)&=h(\delta)+(1-\delta)H(T)+\delta(H(V)+H(W))\nonumber\\
    &=0.65+\frac{5}{6}\cdot 1.571 +\frac{1}{6}\cdot 1 =2.13\ ,
\end{align}
while we have  $H(Y_1,Y_2)=3.04$ when $Y_1$ and $Y_2$ are statistically independent. Similarly, we can compute the entropy of $g(Y_1,Y_2)=U (2T)^{1/2}+(1-U)(V+W)^{1/2}$ as
\begin{align}
\label{joint_graph_entropy_l2_norm}
    H_{G_{Y_1}G_{Y_2}}(Y_1,Y_2)&=h(\delta)+(1-\delta)H((2T)^{1/2})+\delta(H_{G_V}(V)+H_{G_W}(W))\nonumber\\
    &=0.65+(1-\delta)H(T)+\delta(1+0)=2.13\ .
\end{align}
The graph entropy approach   for maximally coupled sources achieves a sum-rate given in~(\ref{joint_graph_entropy_l2_norm}), and hence does not provide an additional gain over the savings of the approach in Section~\ref{section:inner} with a sum-rate achieving (\ref{joint_entropy_l2_norm}). However, for the setting with correlated sources, from (\ref{joint_graph_entropy_l2_norm}) the rate requirement is half of the requirement in (\ref{source_entropy_l2_norm}). 

\begin{figure}[h!]
    \centering
    \includegraphics[width=0.45\columnwidth]{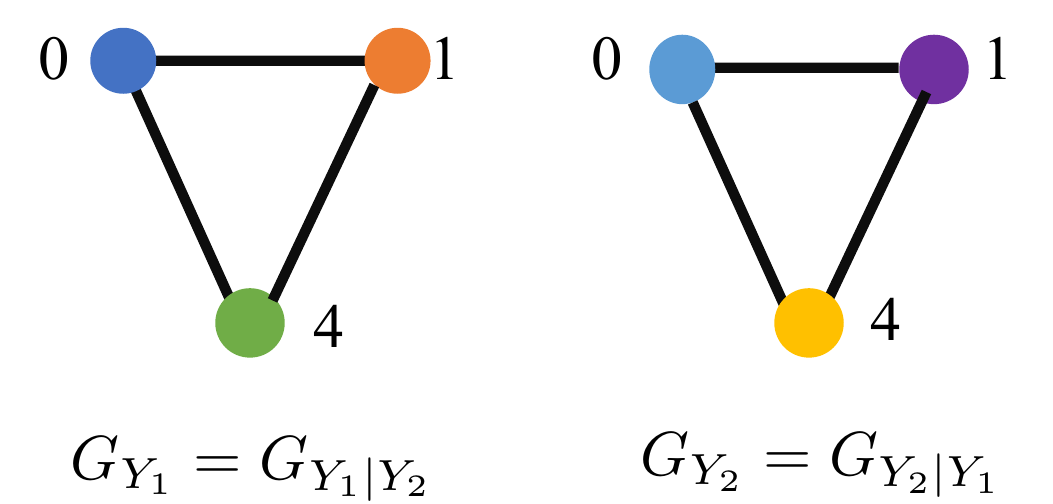}
    \caption{Characteristic graphs for $\ell_2$-norm.}
    \label{fig:p_norm_example}
\end{figure}
\end{enumerate}

\subsection{Computing Polynomials}
\label{section:power2}
We next consider the polynomial of order $m\in\{2k:k\in \mathbb {Z}^+ \}$, i.e., 
\begin{align}
    f(\mathbf X) = \left(\sum_{i=1}^n X_i\right)^m.
\end{align}

We can represent this function via the following decomposition
\begin{align}
     f(\mathbf X)=\Phi\left(\sum_{i=1}^n \psi(X_i)\right)\ , \quad \mbox{where} \quad \Phi(x)=x^m \quad \mbox{and} \quad \psi(x)=x\ .
\end{align}
We assume that sources are uniformly and identically distributed according to ${\cal X}= \{-M,\dots, M\}$. It is easy to note that
\begin{align}
H(\{X_p:p\in\mathcal{S}\})=H(\{Y_p:p\in\mathcal{S}\})=|\mathcal{S}| H(Y_p)=|\mathcal{S}|\log(2M+1)\ .    
\end{align}
Furthermore, $H_{\{G_{Y_p}\}_{p\in\mathcal{S}}}(Y_{\mathcal{S}})<H(\{Y_p:p\in\mathcal{S}\})$ because exploiting the outer functions it is possible to achieve compression gains. For any subset of sources, 
the conditional graph entropies are given by
\begin{align}
&H_{\{G_{Y_p}\}_{p\in\mathcal{S}}}\Big(\{Y_p: p\in\mathcal{S}\}\; \vert \;  \{Y_p: p\notin\mathcal{S}\}\Big)\nonumber\\
&=\sum\limits_{k=-M {|{\cal S}^c}|}^{M {|{\cal S}^c}|}\mathbb{P}\left(\sum\limits_{p\in{\mathcal{S}^c}} X_p=k\right) H_{\{G_{Y_p}\}_{p\in\mathcal{S}}}\Big(\{Y_p: p\in\mathcal{S}\}\,\vert\,\sum\limits_{p\in{\mathcal{S}^c}} X_p=k\Big)\nonumber\\
&\leq|\mathcal{S}| H_{G_{Y_p}}(Y_p)\,, \nonumber    
\end{align}
where the inequality is because $\{G_{Y_p}\}_{p\in\mathcal{S}}$ are jointly distributed. Hence, the graph entropy approach provides savings over the approach in Section~\ref{section:inner}. Since inner functions are identity mappings, computing the inner functions do not provide any additional savings over the compression scheme of Slepian-Wolf.

In order to provide numerical comparisons, consider the following toy example. 
We assume that $X_1,\,X_2\in{\cal X}\triangleq \{-2,-1,0,1,2\}$, and they are uniformly distributed. For this example we do not sketch the characteristic graphs. We note that each distinct outcome of $X_p$ yields a distinct function outcome and, therefore, $H_{G_{X_p}}(X_p)=H(X_p)=\log(M)$. However, the conditional entropies of the characteristic graph of $X_2$ conditional on different possible realizations of $X_1$ satisfy
\begin{align}
    H_{G_{X_2}}(X_2|X_1=x)=2.32 \;\; \mbox{for}\;\;x\in \mathcal{X}\backslash\{0\}\,, \quad \mbox{and} \quad H_{G_{X_2}}(X_2|X_1=0)=h(\frac{2}{5},\frac{2}{5},\frac{1}{5})=1.52\ ,
\end{align}
because $m$ is an even number. This results in a conditional graph entropy $H_{G_{X_2}}(X_2|X_1)=2.16$, which provides savings over the approach in Section~\ref{section:inner}.

\subsection{Computing Maximum and Minimum}
\label{section:maxmin}
We consider the bivariate functions $\max\{X_1,X_2\}$ and  $\min\{X_1,X_2\}$. 
In the simplified KA decomposition form, both of these functions can be represented as
\begin{align}
    f_{\min}(\mathbf X)=\Phi_{\min}((\psi(X_1)+\psi(X_2))\ , \quad \mbox{and} \quad f_{\max}(\mathbf X)=\Phi_{\max}((\psi(X_1)+\psi(X_2))\ ,
\end{align}
where by setting ${\mathbf e}_2 = [0,1,0]^{\top}$ and ${\mathbf e}_3 = [0,0,1]^{\top}$ we have
\begin{align}
 \psi(x)=(1,x,x^2)\ ,    
\end{align}
and 
\begin{align}
\Phi_{\min} (x) & =\frac{x}{2}\cdot {\bf e}_2 - \frac{1}{2}\sqrt{x\cdot 2{\bf e}_3-(x\cdot{\bf e}_2)^2}\ , \\ 
\mbox{and} \quad \Phi_{\max}(x) & =\frac{x}{2}\cdot {\bf e}_2 + \frac{1}{2}\sqrt{x\cdot 2{\bf e}_3-(x\cdot{\bf e}_2)^2}\ .
\end{align}
Note that $\psi(x)$ is a vector-valued function. Hence, the input of the outer function is a vector. To demonstrate the savings of different compression schemes and how they are affected by the source distribution, we consider two examples.

\begin{enumerate}[(i)]
\item {\bf Independent sources.} 
Assume that the $n$ source messages are uniformly and identically distributed over ${\cal X}= \{-M,\dots, M\}$. The entropies of the collection of variables $\{X_{p}: p\in\mathcal{S}\}$ satisfy (\ref{entropy_source_multiplication}) given in~Section~\ref{section:inner}. Note that $H(\phi(X_p))=H({\bf Y}_p)=H(X_p)$. 
When the sources are independently and identically distributed, the graph entropies of ${\bf Y}_p$ for computing $f_{\max}(\mathbf{X})$ and $f_{\min}(\mathbf{X})$ satisfy  $H_{G_{{\bf Y}_{\cal S}}}({\bf Y}_{\cal S}|{\bf Y}_{{\cal S}^c})=|{\cal S}|H(X_p)$ since all outcomes of the sources are to be distinguished. Hence, the graph entropy approach does not provide gain over the scheme in Section \ref{section:inner}. 
For instance, consider the following toy example. Let $X_1\in\{0,1,2,3\}$ and $X_2\in \{0,1\}$ be independent random variables and uniformly distributed. Hence, we have 
\begin{align}
    H(X_1)=2\ , \quad   H(X_2)=1\ , \quad \mbox{and} \quad H(X_1,X_2)=H(X_1)+H(X_2)=3\ .
\end{align}
Note that $f_{\max}(\mathbf{X})\in\{0,1,2,3\}$ with the pmf $\frac{1}{8}(1,3,2,2)$. This yields $H(f_{\max}(\mathbf{X}))=1.9056$. Similarly, $f_{\min}(\mathbf{X})\in\{0,1\}$ with the pmf $\frac{1}{8}(5,3)$, yielding $H(f_{\min}(\mathbf{X}))=0.9544$. We also note that $(X_1+X_2)\in\{0,1,2,3,4\}$ with the pmf $\frac{1}{8} (1,2,2,2,1)$. Hence, its entropy is  $H(X_1+X_2)=2.25$.

Note that $H({\bf Y}_1)=H(X_1)=2$ and $H({\bf Y}_2)=H(X_2)=1$. The characteristic graphs for  $f_{\max}(\mathbf{X})$ and  $f_{\min}(\mathbf{X})$ are shown in Figures \ref{fig:max_example} and \ref{fig:min_example}. The graph entropies of ${\bf Y}_1$ and ${\bf Y}_2$ for computing $f_{\max}(\mathbf{X})$ satisfy  $H_{G_{{\bf Y}_1}}({\bf Y}_1|{\bf Y}_2)=2$, $H_{G_{{\bf Y}_2}}({\bf Y}_2)=1$ and $H_{G_{{\bf Y}_1},G_{{\bf Y}_2}}({\bf Y}_1,{\bf Y}_2)=3$ since all outcomes of both sources are to be distinguished. This yields the following relationship:
\begin{align}
H_{G_{{\bf Y}_1},G_{{\bf Y}_2}}({\bf Y}_1,{\bf Y}_2)=3&=H({\bf Y}_1,{\bf Y}_2) \\
&>H(X_1+X_2)=2.25>H(f_{\max}(\mathbf{X}))=1.9056\ .
\end{align}
Hence, the graph entropy approach does not provide gain over the scheme in Section \ref{section:inner}, which requires a sum rate $H({\bf Y}_1,{\bf Y}_2)=3$.

The graph entropies of ${\bf Y}_1$ and ${\bf Y}_2$ for computing $f_{\min}(\mathbf{X})$ satisfy  $H_{G_{{\bf Y}_1}}({\bf Y}_1|{\bf Y}_2)=0.5$ because $X_1$ only needs to distinguish between the sets $\{0\}$ and $\{1\}$ and $X_1\in \{2,3\}$ does not cause any confusion at the receiver, and $H_{G_{{\bf Y}_2}}({\bf Y}_2)=1$ and $H_{G_{{\bf Y}_1},G_{{\bf Y}_2}}({\bf Y}_1,{\bf Y}_2)=1.5$. Hence, the rates satisfy:
\begin{align}
H_{G_{{\bf Y}_1},G_{{\bf Y}_2}}({\bf Y}_1,{\bf Y}_2)=1.5>H(f_{\min}(\mathbf{X}))=0.95\ .    
\end{align}
In this scheme, the graph entropy approach requires half of the rate of the scheme in Section~\ref{section:inner}, which requires a sum rate $H({\bf Y}_1,{\bf Y}_2)=3$. 

\begin{figure}[t!]
    \centering
    \includegraphics[width=0.9\columnwidth]{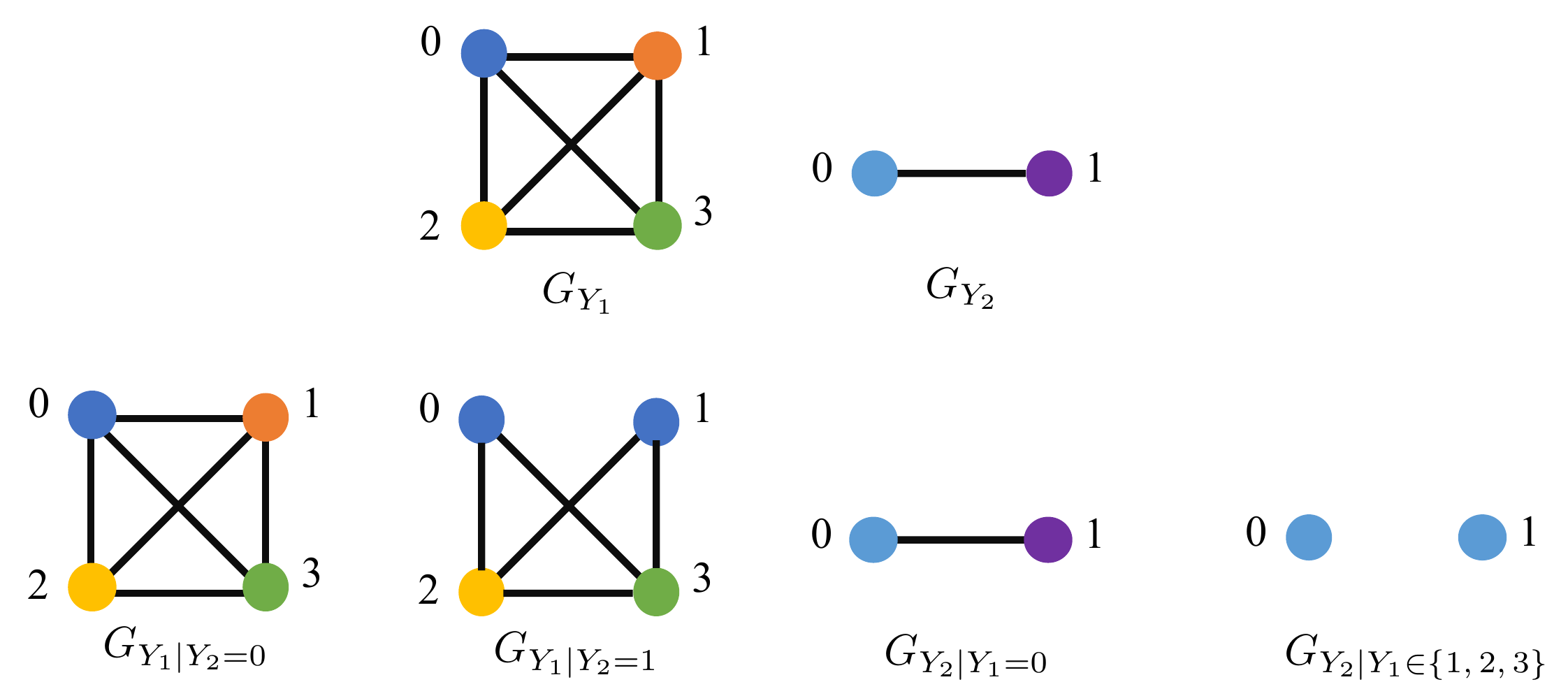}
    \caption{Characteristic graphs for $f_{\max}(\mathbf{X})$.}
    \label{fig:max_example}
\end{figure}

\begin{figure}[t!]
    \centering
    \includegraphics[width=0.9\columnwidth]{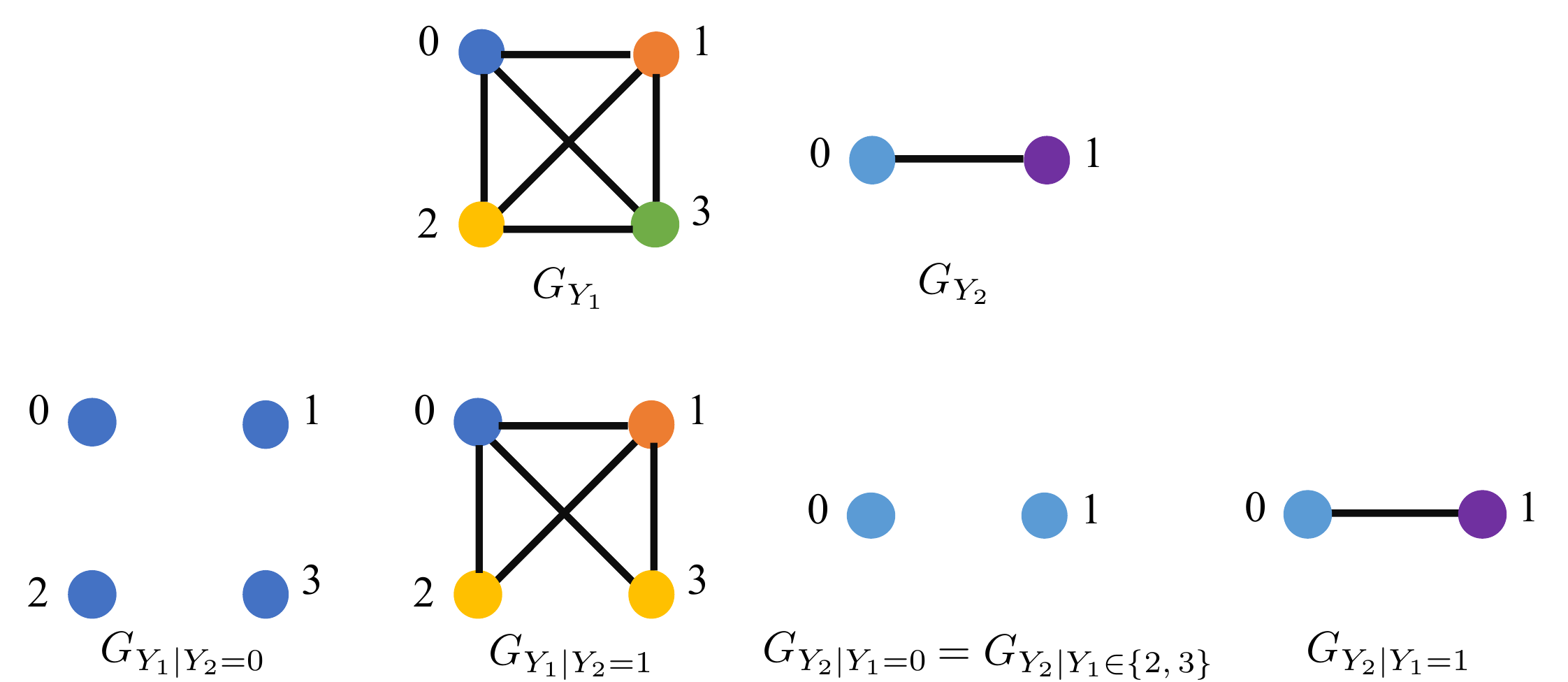}
    \caption{Characteristic graphs for $f_{\min}(\mathbf{X})$.}
    \label{fig:min_example}
\end{figure}

\item {\bf Correlated sources.} We let $X_1\in \{0,1,4\}$ and $X_2\in\{0,1,4\}$ have marginal pmfs $\frac{1}{3}(1,1,1)$ and $\frac{1}{4}(2,1,1)$, respectively. We also let $Y_1$ and $Y_2$ to have the same marginal pmfs as $X_1$ and $X_2$, respectively and be statistically dependent. Furthermore, we assume that $\mathbb{P}(Y_1=Y_2)$ takes its maximum value, which is achieved via maximal coupling. In other words, we can rewrite $Y_1$ and $Y_2$ according to \cite{grimmett2001probability}:
\begin{align}
    Y_1=UT+(1-U)V\ , \quad \mbox{and} \quad    Y_2=UT+(1-U)W\ ,
\end{align}
where $U\sim\mbox{Bern}(1-\delta)$ and $\delta=\frac{1}{6}$, and the discrete random variables $T,\,V$, and $W$ have the pmfs $p_T=(\frac{2}{5},\frac{3}{10},\frac{3}{10})$, $p_V=\frac{1}{2}(0,1,1)$, and $p_W=(1,0,0)$  at the respective coordinates, respectively. Hence, their entropies are $H(T)=h(\frac{2}{5},\frac{3}{10},\frac{3}{10})=1.571$, $H(V)=1$, and $H(W)=0$. In this setting the joint entropy of the variables is given by
\begin{align}
    H(Y_1,Y_2)&=h(\delta)+(1-\delta)H(T)+\delta (H(V)+H(W))\nonumber\\
    &=0.65+\frac{5}{6}\cdot 1.571 +\frac{1}{6}\cdot 1=2.13\ ,\nonumber
\end{align}
while we have $H(X_1,X_2)=3.085$ when $X_1$ and $X_2$ are independent.

Similarly, we can compute the entropy of $f_{\max}(\mathbf{X})=UT+(1-U)\max(V,W)$ as
\begin{align}
H(f_{\max}(\mathbf{X})))&=h(\delta)+\frac{5}{6}H(T)+\frac{1}{6}H(f_{\max}(V,W)) \\
&=0.65+\frac{5}{6}H(T)+\frac{1}{6}H(V)=2.13\ .
\end{align}

We note that the conditional graph entropies of ${\bf Y}_1$ and ${\bf Y}_2$ for computing $f_{\max}(\mathbf{X})$ can be determined by eliminating the common parts and encoding the independent parts $V$ and $W$. We note that $H_{G_{{\bf Y}_1}}({\bf Y}_1)=H(X_1)=2.13$ and $H_{G_{{\bf Y}_2}}({\bf Y}_2)=H(X_2)=1.96$ because $X_1$ and $X_2$ need to distinguish between all outcomes not to cause any confusion at the receiver. Furthermore, the sources satisfy  $H_{G_{{\bf Y}_1}}({\bf Y}_1\vert {\bf Y}_2)=\frac{1}{6}H_{G_{V}}(V)=\frac{1}{6}$, and similarly, $H_{G_{{\bf Y}_2}}({\bf Y}_2\vert {\bf Y}_1)=\frac{1}{6}H_{G_{W}}(W)=0$ and $H_{G_{{\bf Y}_1}G_{{\bf Y}_2}}({\bf Y}_1,{\bf Y}_2)=2.13$.

The entropy of $f_{\min}(\mathbf{X}))=UT+(1-U)\min(V,W)$ is given by
\begin{align}
H(f_{\min}(\mathbf{X})))&=h(\delta)+\frac{5}{6}H(T)+\frac{1}{6}H(f_{\min}(V,W))\nonumber\\
&=0.65+\frac{5}{6}H(T)+\frac{1}{6}H(W)=1.96\ .\nonumber
\end{align}

Due to the symmetry of the source distributions, the graph entropies of ${\bf Y}_1$ and ${\bf Y}_2$ for computing $f_{\min}(\mathbf{X})$ are identical to the graph entropies for computing $f_{\max}(\mathbf{X})$. We have the following comparison between the entropies:
\begin{align}
H_{G_{{\bf Y}_1}G_{{\bf Y}_2}}({\bf Y}_1,{\bf Y}_2)=H(X_1,X_2)=H(f_{\max}(\mathbf{X}))=2.13>H(f_{\min}(\mathbf{X})=1.96\ . \nonumber 
\end{align}

\end{enumerate}

From this example we see that when  $Y_1$ and $Y_2$ are maximally coupled, the conditional graph entropies for computing $f_{\min}(\mathbf{X})$ and $f_{\max}(\mathbf{X})$ can be reduced significantly, resulting a significant reduction in the sum-rate required for computing and approximating the entropy of the functions.

\subsection{Affine Transformations}
\label{section:affine}

We consider the affine function in the form 
\begin{align}
f(\mathbf{X})=\sum\limits_{p=1}^n \alpha_p X_p\ ,\quad \alpha_p\in\mathbb{R}\ .    
\end{align}

Leveraging the decomposition of $f(\mathbf{X})$ in~\eqref{single_inner_representation}, this can be represented as
\begin{align}
    f(\mathbf X)=\Phi\left(\sum\limits_{p=1}^n\alpha_p\psi(X_p)\right)\ , \quad \mbox{where} \quad \Phi(x)=x \quad \mbox{and} \quad \psi(x)=x\ .
\end{align}
Hence both the inner and the outer functions are identity mappings. 

To demonstrate the savings of different compression schemes and how they are affected by the source distribution, we consider discrete-valued source variables. For discrete source variables, it is easy to note that $H_{G_{X_p}}(X_p)=H(X_p)$ for $p\in \{1,\dots,n\}$ because  $G_{X_p}$'s are complete graphs. The saving in compression is determined by $H_{G_{X_{\cal S}}}(X_{\cal S}|X_{{\cal S}^c})$. If $\{X_p\}_p$ are independent, it is not possible to overperform the compression scheme of Slepian-Wolf. When the variables are dependent, note that 
\begin{align}
H_{\{G_{X_p}\}_{p\in{\cal S}}} (X_{\cal S}| X_{{\cal S}^c})&=   \min_{\overset{\overset{X_p\in W_p\in \Gamma(G_{X_p})}{ W_p-X_p-X_r}}{p\in{\cal S},\, r\in {\cal S}^c}} I\Big(\{W_p: p\in\mathcal{S}\}; \{X_p: p\in\mathcal{S}\}\vert  \{X_r: r\in\mathcal{S}^c\}\Big) \nonumber\\
&=H(X_{\cal S}| X_{{\cal S}^c})-\min_{\overset{\overset{X_p\in W_p\in \Gamma(G_{X_p})}{ W_p-X_p-X_r}}{p\in{\cal S},\, r\in {\cal S}^c}} H\Big(X_{\cal S}| X_{{\cal S}^c},\, \{W_p: p\in\mathcal{S}\}\Big)\nonumber\\
&=H\Big(X_{\cal S}| X_{{\cal S}^c}\Big)\ ,\nonumber
\end{align}
where in the second step we used that for any $x_p$ the set $\{x_{\cal S}: p(x_{\cal S}|x_{{\cal S}^c})>0,\, f(\mathbf x)=\sum\limits_{p=1}^n\alpha_p x_p\}$ of possible $x_{\cal S}$ forms a clique, and $\{W_p:p\in\mathcal{S}\}$ fully specifies $X_{\cal S}$ given $X_{{\cal S}^c}$. Hence, the savings of graph coloring approach may not outperform that of Slepian-Wolf.

In order to highlight the key ideas, consider the simple setting of $n=2$.
The saving in compression is determined by $H_{G_{X_2}}(X_2|X_1)$. If $X_1$ and $X_2$ are independent it is not possible to overperform the compression scheme of Slepian-Wolf. When the variables are dependent, note that 
\begin{align}
H_{G_{X_2}}(X_2|X_1)&=\min_{\overset{X_2\in W_2\in \Gamma(G_{X_2})}{ W_2-X_2-X_1}} I(W_2;X_2\vert X_1)\nonumber\\
&= H(X_2|X_1)-\max_{X_2\in W_2\in \Gamma(G_{X_2})}H(X_2|X_1,W_2)    \nonumber\\
& =H(X_2|X_1)\ ,\nonumber
\end{align}
where in the second step we used that for any $x_1$ the set $\{x_2: p(x_2|x_1)>0,\, f(x_1,x_2)=\alpha_1 x_1+\alpha_2 x_2\}$ of possible $x_2$ forms a clique, and $W_2$ fully specifies $X_2$ given $X_1$. Hence, the savings of graph coloring approach may not outperform that of Slepian-Wolf.


\subsection{Computing Gradient}
\label{section:gradient}

Computing gradients arises in a wide range of distributed multi-agent optimization settings, in which optimizing a utility is carried out by computing local counterparts of the utility by the agents followed by exchanging local gradient values. In this subsection, we aim to compute the gradient of a generic $f(x)$ whose associated inner functions $\psi$ are differentiable. Based on the representation of $f(x)$ in (\ref{multiple_inner_representation}) we have
\begin{align}
\label{gradient1}
\frac{\partial f(\mathbf{X})}{\partial X_p} 
 &=\frac{\partial}{\partial X_p}\sum\limits_{q=0}^{2n} \Phi_q \left(\sum\limits_{p=1}^n \psi_{q,p} (X_p) \right) \\
&=\sum\limits_{q=0}^{2n}\Phi'_q\left(\sum\limits_{p=1}^n \psi_{q,p} (X_p)\right) \cdot \frac{d}{d X_p}\left(\sum\limits_{p=1}^n \psi_{q,p}(X_p)\right)\\
\label{gradient3} & = \sum\limits_{q=0}^{2n} \Phi'_q\left(\sum\limits_{p=1}^n \psi_{q,p} (X_p)\right) \cdot \frac{d \psi_{q,p}(X_p)}{d X_p}\ .   
\end{align}
It is noteworthy that the partial derivative of $f(\mathbf{X})$ with respect to $X_p$ only modifies the outer function, which also depends on the $p^{\rm th}$ coordinate of $\mathbf{X}$. If we use the Sprecher's representation in~\eqref{single_inner_representation} we have $Y_{pq}= \alpha_p \psi(X_p+qa)$, and $Y_q=\sum\limits_{p=1}^n Y_{pq}$,  and we have 
\begin{align}
\nabla f(\mathbf{X}) = \sum\limits_{q=0}^{2n} \Phi'\left(Y_q+q\right)\cdot \nabla Y_{q}\ .
\end{align}
This will only change the post-decoding model in Figure \ref{fig:innercoded}. Specifically, instead of computing $\Phi(Y_q+q)$, we compute $\Phi'\left(Y_q+q\right)\cdot \nabla Y_q$.

\begin{figure}[h!]
    \centering
    \includegraphics[width=\columnwidth]{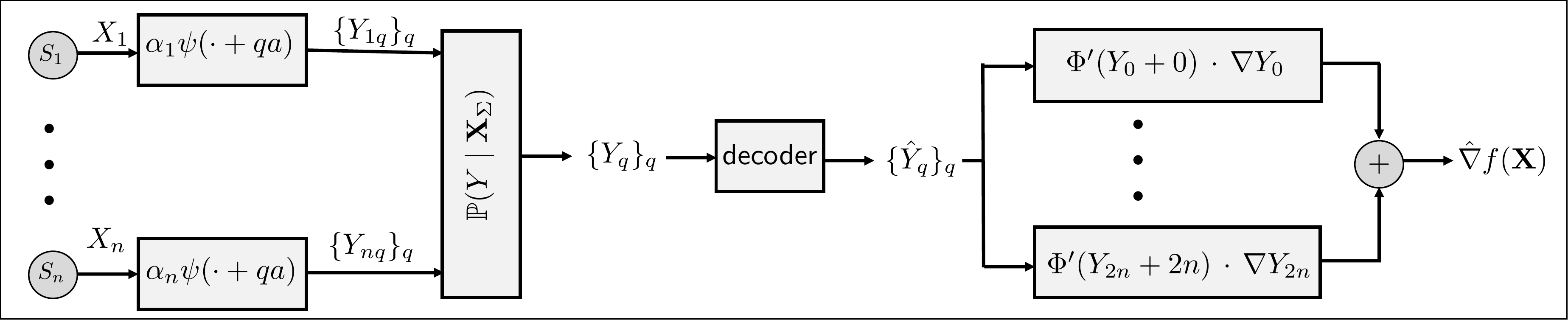}
    \caption{Transmitting inner function representations for computing the gradient.}
    \label{fig:innercoded_gradient}
\end{figure}

\subsection{Computing Hessian}
\label{section:hessian}

Building on gradient computation, when $f(\mathbf{X})$ is twice-differentiable with respect to all coordinate pairs, we also compute its Hessian matrix. The Hessian matrix of $f(\mathbf{X})$ is a square $n\times n$ matrix given by
\begin{align}
\mathbf{H}(f(\mathbf{X}))=
\begin{bmatrix}    
\frac{\partial^2 f(\mathbf{X})}{\partial X_1^2} & \frac{\partial^2 f(\mathbf{X})}{\partial X_1\,\partial X_2} & \dots & \frac{\partial^2 f(\mathbf{X})}{\partial X_1\,\partial X_n}\\
\frac{\partial^2 f(\mathbf{X})}{\partial X_2\,\partial X_1} & \frac{\partial^2 f(\mathbf{X})}{\partial X_2^2} & \dots & \frac{\partial^2 f(\mathbf{X})}{\partial X_2\,\partial X_n}\\
\vdots & \vdots & \ddots & \vdots \\
\frac{\partial^2 f(\mathbf{X})}{\partial X_n\,\partial X_1} & \frac{\partial^2 f(\mathbf{X})}{\partial X_n\,\partial X_2} & \dots & \frac{\partial^2 f(\mathbf{X})}{\partial X_n^2}
\end{bmatrix}\ .
\end{align}
For the $(p,\,r)$ array of  $\mathbf{H}(f(\mathbf{X}))$ we then have
\begin{align}
\label{hessian}
(\mathbf{H}(f(\mathbf{X})))_{p,\,r}
&=\sum\limits_{q=0}^{2n} \Phi''_q\left(\sum\limits_{p=1}^n \psi_{q,p} (X_p)\right)\cdot \frac{d \psi_{q,p}(X_p)}{d X_p} \cdot \frac{d \psi_{q,r}(X_r)}{d X_r}\ .
\end{align}
Based on Sprecher's representation we have $Y_{pq}= \alpha_p \psi(X_p+qa)$, and $Y_q=\sum\limits_{p=1}^n Y_{pq}$, based on which (\ref{hessian}) simplifies to 
\begin{align}
(\mathbf{H}(f(\mathbf{X})))_{p,\,r}=\sum\limits_{q=0}^{2n}  \Phi''_q\left(Y_q\right)\cdot \frac{d Y_q}{d X_p}\cdot \frac{d Y_q}{d X_r}\ ,
\end{align}
where 
\begin{align}
\frac{d Y_q}{d X_p}=\frac{d Y_{pq}}{d X_p} \ , \qquad \mbox{and} \quad \frac{d Y_q}{d X_r}=\frac{d Y_{rq}}{d X_r}\ .
\end{align}
Furthermore, by setting $\Phi_q(Y_q)=\Phi(Y_q+q)$ we have
\begin{align}
\frac{\partial^2 f(\mathbf{X})}{\partial X_p\,\partial X_r}=    \sum\limits_{q=0}^{2n} \Phi''\left(Y_q+q\right)\cdot \frac{d Y_{pq}}{d X_p}\cdot \frac{d Y_{rq}}{d X_r}\ .
\end{align}
Hence, the Hessian matrix of $f(\mathbf{X})$ is
\begin{align}
\mathbf{H}(f(\mathbf{X}))= \sum\limits_{q=0}^{2n} \Phi''\left(Y_q+q\right)\cdot \nabla Y_{q} \cdot \nabla^{\top} Y_{q}\ ,
\end{align}
where $\nabla^{\top}$ denotes the transpose of $\nabla$ operator.

\begin{figure}[h!]
    \centering
    \includegraphics[width=\columnwidth]{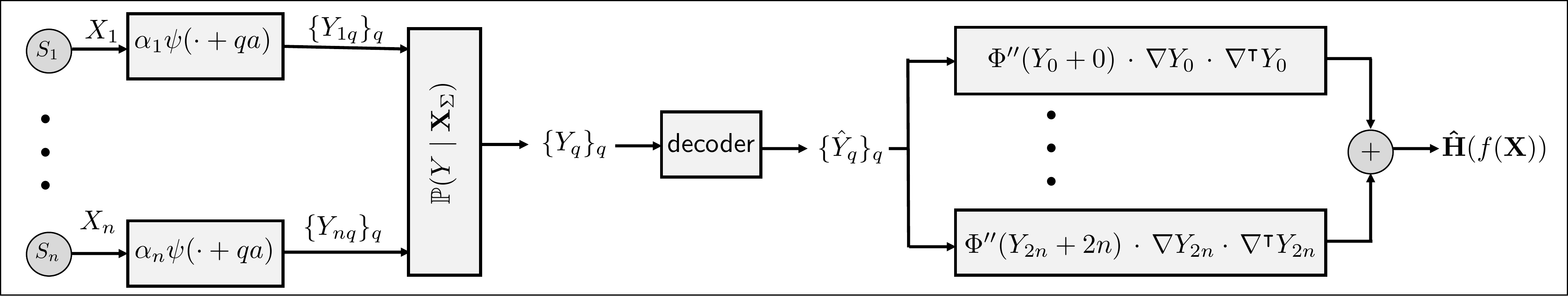}
    \caption{Transmitting inner function representations for computing the Hessian.}
    \label{fig:innercoded_hessian}
\end{figure}

By following the same line of analyses as in sections~\ref{section:gradient} and \ref{section:hessian}, we can compute higher-order differential values. In the higher orders, similarly to the first and second orders, the inner functions remain the same, and the effect of the higher degree derivatives of $\Phi(Y_q+q)$ and $Y_q$ will appear in the outer functions.

\subsection{Computing Real Analytic Functions}
\label{section:analytic}

\begin{figure}[t!]
    \centering
    \includegraphics[width=\columnwidth]{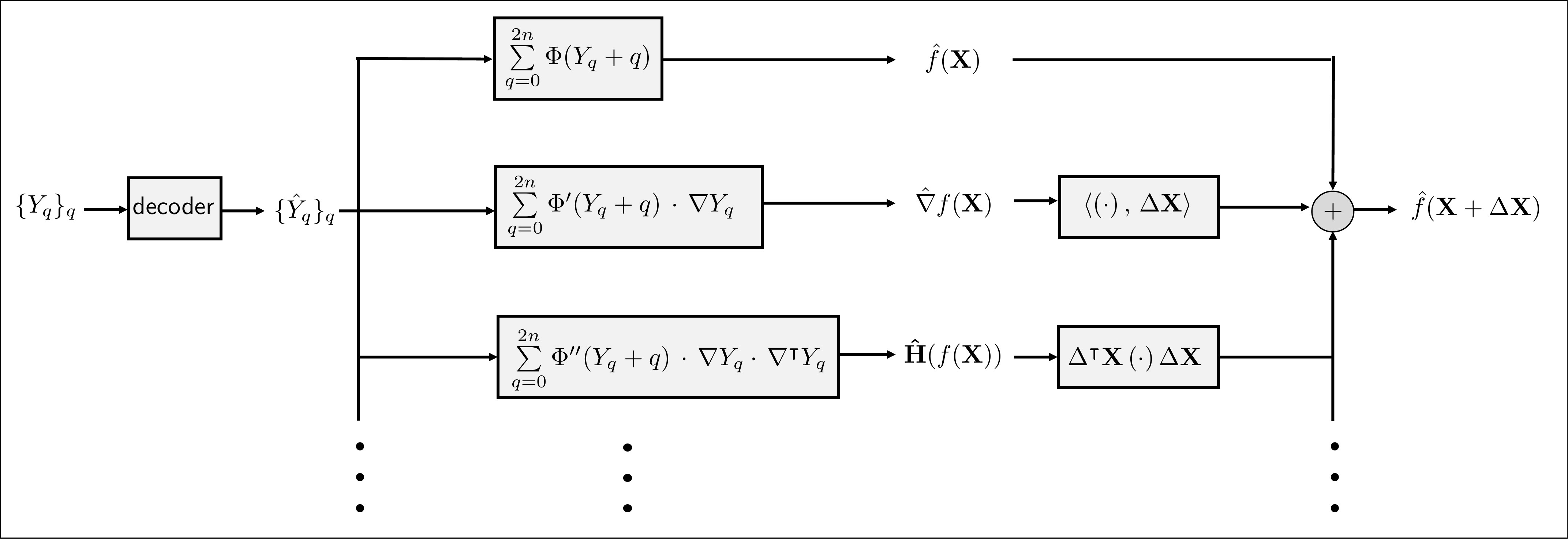}
    \caption{Computing the Taylor series approximation.}
    \label{fig:innercoded_taylor}
\end{figure}

By leveraging the results and discussions in sections~\ref{section:gradient} and \ref{section:hessian}, next, we provide a framework for computing generic real analytic functions that are infinitely differentiable. In particular, we focus on functions for which the Taylor series at any point $\mathbf{X}_0$ in its domain exhibits pointwise convergence to $f(\mathbf{X})$ for $\mathbf{X}$ in a neighborhood of $\mathbf{X}_0$. For a function $f(\mathbf{X})$ satisfying these, the Taylor expansion is given by 
\begin{align}
f(\mathbf{X})=\sum\limits_{|i|=0}^{\infty} \frac{D^{i }f(\mathbf{X}_0)}{i!}(\mathbf{X}-\mathbf{X}_0)^i    \ ,
\end{align}
where 
\begin{align}
    D^{i }f \triangleq {\frac {\partial ^{|i |}f}{\partial x_{1}^{i _{1}}\cdots \partial x_{n}^{i _{n}}}}\ ,\qquad |i |\geq 0\ ,
\end{align}
and for $i\in\mathbb{N}^n$ we have defined
\begin{align}
|i |=i _{1}+\cdots +i _{n}\ ,\quad i !=i _{1}!\cdots i _{n}!\ ,\quad \mbox{and} \quad  {\mathbf{X}}^{i }=X_{1}^{i _{1}}\cdots X_{n}^{i _{n}}\ .
\end{align}
For instance, for the second-order approximation of $f(\mathbf{X})$ we have
\begin{align}
f(\mathbf{X}+\Delta\mathbf{X})&\approx f(\mathbf{X})+ \nabla^{\top} f(\mathbf{X})\Delta\mathbf{X}+\frac{1}{2} \Delta\mathbf{X}^{\top} \mathbf{H}(f(\mathbf{X})) \Delta\mathbf{X}\nonumber\\
&=f(\mathbf{X})+ \left(\sum\limits_{q=0}^{2n} \Phi'\left(Y_q+q\right)\cdot \nabla Y_{q}\right)^{\top}\Delta\mathbf{X}\nonumber\\
&+\frac{1}{2} \Delta\mathbf{X}^{\top} \left(\sum\limits_{q=0}^{2n} \Phi''\left(Y_q+q\right)\cdot \nabla Y_{q} \cdot \nabla^{\top} Y_{q}\right)\Delta\mathbf{X}\ .
\end{align}
Hence, for the local expansion of the gradient we have
\begin{align}
\nabla f(\mathbf{X}+\Delta\mathbf{X})= \nabla f(\mathbf{X})+  \mathbf{H}(f(\mathbf{X})) \Delta\mathbf{X}+O(\norm{\Delta\mathbf{X}}^2)\ .
\end{align}
Hence, instead of explicitly computing 
\begin{align}
f(\mathbf{X}+\Delta\mathbf{X})=\sum\limits_{q=0}^{2n} \Phi_{q,p} \left(\sum\limits_{p=1}^n \psi_{q,p} (X_p+\Delta X_p) \right) \ , 
\end{align}
we only need to change the post-processing blocks as shown in Figure~\ref{fig:innercoded_taylor}.

\section{Conclusions \& Extensions}
\label{section:extensions}

We have exploited Kolmogorov's superposition theorem to devise a decomposition-based framework for distributed computation of functions over the additive MAC. The framework is modular since it provides a two-layer nested superposition of inner and outer function computations conducted at the sources or the receiver, and is universal since the inner functions are independent of the function of interest. It also leverages the additive multiple access channels (MACs) for the superposition of the transmissions. Our framework provides savings by compressing the inner functions at the sources and computing the outer function at the receiver.  Extra savings are possible by leveraging a graph entropy perspective to partially compress the outer functions at the source sites. This approach enables the flexibility of distributing computation across the source and receiver sites. In practical settings, the sources and receivers might have different computation capabilities. Hence, the proposed technique can provide insights into effective distributed computation of tasks over the network accounting for the bottlenecks.

In the proposed framework, having feedback can improve the quality of the functional representation. Extensions of this work include investigating more general multi-terminal frameworks beyond MAC and exploiting the structural match between the channel and the observations, possibly with side information and feedback, and recovering approximations instead of exact representations (via adapting the inner and the outer functions) subject to a distortion criterion, and also modeling continuous sources and their graph colorings via quantization. Further generalizations include computation under privacy constraints and the multi-functional problem where the receiver aims to compute some deterministic functions, exploiting multi-functional graph entropy.


\bibliographystyle{IEEEtran}
\bibliography{ref}
\IEEEtriggeratref{3}

\end{document}